\documentclass[%
reprint,
superscriptaddress,
amsmath,amssymb,
pra, 
floatfix,
nobalancelastpage
]{revtex4-2}

\usepackage[dvipsnames]{xcolor}
\usepackage[colorlinks]{hyperref}
\hypersetup{
linkcolor=BrickRed
,citecolor=NavyBlue 
,filecolor=Mulberry
,urlcolor=NavyBlue
,menucolor=BrickRed
,runcolor=Mulberry
,linkbordercolor=BrickRed
,citebordercolor=Green
,filebordercolor=Mulberry
,urlbordercolor=NavyBlue
,menubordercolor=BrickRed
,runbordercolor=Mulberry
}
\usepackage{nicefrac}
\usepackage{float}
\usepackage{amsmath}
\usepackage{amssymb}
\usepackage{wasysym}
\usepackage{multirow}
\usepackage[normalem]{ulem}
\usepackage{upgreek}
\usepackage{braket}
\usepackage{siunitx}

\usepackage{graphicx}
\usepackage{dcolumn}
\usepackage{bm}

\def\hyperef#1#2{\hyperref[#2]{#1~\ref{#2}}}

\def\lf{\left\lfloor}   
\def\rf{\right\rfloor}

\newcommand{\fLO}{f_\text{LO}}

\begin{document}

\title{Quantum Limits of Superconducting-Photonic Links \& Their Extension to mm-Waves}
\author{Kevin K. S. Multani}
\thanks{These authors contributed equally to this work.}
\affiliation{%
Department of Physics, Stanford University, Stanford, California 94305.
}
\affiliation{%
 Department of Applied Physics and Ginzton Laboratory, Stanford University, Stanford, California 94305.
}
\author{Wentao Jiang}%
\thanks{These authors contributed equally to this work.}
\affiliation{%
 Department of Applied Physics and Ginzton Laboratory, Stanford University, Stanford, California 94305.
}
\author{Emilio A. Nanni}
\affiliation{%
SLAC National Accelerator Laboratory, Stanford University, Menlo Park, California 94025.
}
 \author{Amir H. Safavi-Naeini}
 \email{safavi@stanford.edu}
\affiliation{%
 Department of Applied Physics and Ginzton Laboratory, Stanford University, Stanford, California 94305.
}

\date{\today}

\begin{abstract}
Photonic addressing of superconducting circuits has been proposed to overcome wiring complexity and heat load challenges, but superconducting-photonic links suffer from an efficiency-noise trade-off that limits scalability. This trade-off arises because increasing power conversion efficiency requires reducing optical power, which makes the converted signal susceptible to shot noise. We analyze this trade-off and find the infidelity of qubit gates driven by photonic signals scales inversely with the number of photons used, and therefore the power efficiency of the converter. While methods like nonlinear detection or squeezed light could mitigate this effect, we consider generating higher frequency electrical signals, such as millimeter-waves (100 GHz), using laser light. 
At these higher frequencies, circuits have higher operating temperatures and cooling power budgets. We demonstrate an optically-driven cryogenic millimeter-wave source with a power efficiency of $10^{-4}$ that can generate \SI{1}{\micro\watt} of RF power at 80 GHz with 1500 thermal photons of added noise at $\SI{4}{\K}$. Using this source, we perform frequency-domain spectroscopy of superconducting NbTiN resonators at 80-90 GHz. Our results show a promising approach to alleviate the efficiency-noise constraints on optically-driven superconducting circuits while leveraging the benefits of photonic signal delivery. Further optimization of power efficiency and noise at high frequencies could make photonic control of superconducting qubits viable at temperatures exceeding $\SI{1}{\K}$.
\end{abstract}

\maketitle

\section{\label{sec:intro}Introduction}
The field of superconducting quantum computing needs efficient and scalable methods to control and read out qubits. Current approaches using attenuated coaxial lines face limitations regarding heat load and wiring complexity as the number of qubits scales toward the millions needed for fault-tolerant quantum computing. Optically-driven microwave devices present an attractive alternative, leveraging the low thermal conductivity and large bandwidth of optical fibers to deliver many control signals with minimal heat load.

In classical applications, optically-driven techniques for RF signal generation, distribution, and processing, fall under the umbrella of  microwave photonics, and have been actively investigated~\cite{capmany2007microwave, payne2002photonic, bagnell2013millimeter, Niu2015MoF, bardalen2018evaluation, burla2019500, miller2017attojoule, marpaung2019integrated, zhu2023integrated, kudelin2024photonic, sun2024integrated}. In turn, the photonic approach for signal distribution was adapted for quantum control and readout of a superconducting qubit operating at $\SI{5}{\GHz}$~\cite{lecocq2021control, li2024optical}. 

We begin by evaluating the potential of using modulated lasers and cryogenic photodetection to create microwave drive fields within a dilution refrigerator for superconducting qubit systems. The current leading approach, in terms of power efficiency $\eta_\text{P} = P_{\mu\text{w}}/P_0$ (the ratio between microwave output power and optical input power), uses photodiodes to detect modulated laser light. Given the small heat budget available at cryogenic temperatures, we find that the power efficiency must be improved over the current state of the art to allow for many drive lines. While considering the consequences of improving the power efficiency, we encounter a fundamental limitation stemming from optical shot-noise, which makes implementing such approaches while maintaining low noise performance technically challenging.

As an alternative, we consider higher frequency microwave systems which can operate at higher temperatures. We find that the equivalent temperature of optically generated microwave radiation decreases as we go to higher frequencies. Moreover, by operating at higher temperatures where a significantly larger heat budget is possible, the issues related to shot-noise also become more easily surmountable, since we can reduce the efficiency of the optical-electric conversion. Turning our attention to the mm-wave regime, where progress is being made in superconducting qubit development~\cite{faramarzi2021initial,anferov2024improved,anferov2024superconducting}, we construct and characterize an optically-driven cryogenic millimeter-wave source (mm-wave or $\Omega/2\pi \approx100~\mathrm{GHz}$). We measure our source's power efficiency and noise properties and use it to probe superconducting mm-wave circuits~\cite{stokowski2019towards, multani2020development, anferov2020millimeter}. Our source consists of one diode laser in the optical C-band (1550 nm), a high-bandwidth electro-optic modulator, and a high-speed photodiode. The photodiode is mounted on the 4 K stage of a dilution refrigerator, where the mm-waves are generated~\cite{Zhang1997cryogenic, huggard2007photonic, bardalen2018evaluation}.

\section{The challenge with optical-to-microwave conversion}
Suppose we are trying to generate microwave pulses to drive the gate lines of $10^4$ superconducting qubits~\cite{krantz2019quantumeng} using laser light. For simplicity, we use one photodiode per drive line, then send an optical signal to each photodiode to generate a voltage or current signal that drives each qubit. Our goal is to dissipate less than $10^{-5}$ watts of power in total, which means that each of photodiode will need to absorb about $10^{-9}$ watts of optical power on average. Qubit gates in this architecture are implemented by electrical pulses with duration on the order of $10^{-8}$ seconds. This means that each gate pulse will need to be encoded with an optical energy of roughly $10^{-17}$ joules, which corresponds to roughly $10^{2}$ optical photons. Optical shot noise would then lead to fluctuations in energy from gate to gate of roughly $10\%$. As we will show below, this leads to a gate infidelity on the order of a few percent. In other words, quantum fluctuations in the light field fundamentally limit approaches for optically-driven microwave control of superconducting systems.

In the following sections, we consider the above example in more detail, and consider strategies for improving the prospects for this approach.

\subsection{Optical-to-Electrical Power Conversion Efficiency}
\label{subsec:figuresofmerit}
To understand the performance of optically driven microwave sources, we consider a system where an optical signal is intensity modulated at the desired frequency $\Omega$ and sent onto a photodiode to generate a microwave signal. The optical power is given by,
\begin{equation}
    P_{\text{opt}}(t) = P_0\Big(1+\epsilon_\text{m} \cos(\Omega t)\Big),
\end{equation}
where $P_0$ is the time-averaged optical power, $\epsilon_\text{m}$ is the modulation depth ($|\epsilon_\text{m}|\le1$), and $\Omega$ is the modulation frequency.

When light at a sufficiently high frequency $\omega_\mathrm{opt}$ impinges on a photodiode, carriers are generated by absorption of photons. The photodiode converts this optical signal into an electrical current proportional to the instantaneous optical power. If all the incident photons are detected by the generation of a carrier of charge $e_0$, then we achieve a responsivity,
\begin{equation}
\mathcal{R}_\text{q} = \frac{e_0}{\hbar \omega_\text{opt}}  \approx~\SI{1.25}{\A\per\watt} \text{ at } \lambda = 1550~\text{nm},
\end{equation}
where $e_0$ is the electron charge, $\hbar$ is the reduced Planck constant, and $\omega_\text{opt}$ is the optical frequency. The observed responsivity may be modified by the quantum efficiency ($\eta_\text{q}$) and gain of internal or external amplification processes ($G$), giving 
\begin{equation}
\mathcal{R} = G\eta_\text{q}\mathcal{R}_\text{q}.    
\end{equation}
The instantaneous photocurrent generated by the photodiode is given by $I(t) = \mathcal{R}P_{\text{opt}}(t)$. When this photocurrent is sent into a microwave transmission line with characteristic impedance $Z_0$, the resulting average microwave power at the modulation frequency $\Omega$ is,
\begin{equation}
P_{\mu\text{w}} = \frac{1}{2}\mathcal{R}^2 P_0^2 \epsilon_\text{m}^2 Z_0.
\end{equation}
The power efficiency, defined as the ratio of microwave power output to optical power input, is then,
\begin{equation}
\eta_\text{P} = \frac{P_{\mu\text{w}}}{P_0} = \frac{1}{2} \mathcal{R}^2  \epsilon_\text{m}^2 Z_0 P_0.\label{eqn:eta_p}
\end{equation} 
The maximum number of drive lines, and thus, qubits is limited by cooling power of the environment and can be written as,
\begin{equation}
    N_\text{qubit} = \lf \frac{P_\text{cool}}{P_0^\text{op}} \rf = \lf\eta_\text{P} \left(\frac{P_\text{cool}}{P_{\mu\text{w}}^{\text{op}}}\right)\rf,\label{eqn:Nqubit}
\end{equation}
where $P_\text{cool}$ is the cooling power of the environment that the qubits are thermalized to, $P_0^\text{op}$ is the optical power required to perform a qubit operation (state preparation, gate operation, etc.), and $P_{\mu \text{w}}^\text{op}$ is the corresponding microwave power. Considering the $10^{-6}$ watts of cooling power available in the state-of-the-art dilution refrigerators, the analysis above suggests that at most $10^2$ qubits can have their gate lines optically-driven with unamplified photodiodes before deleterious effects of heating begin to dominate. This limit highlights the need for improved power efficiency to scale to the larger numbers of physical qubits that fault-tolerant quantum computing requires.

\subsection{Shot-Noise in Optically Generated Electrical Signals}
\label{sec:effectivetemp}
Noise and fluctuations in the optically-generated electrical signals may strongly affect the quantum systems we intend to control. A fundamental source of noise is the shot-noise of the detected optical field, which we encapsulate using an effective photon noise occupation at the microwave frequency, $n_\text{eff}(\Omega)$. This effective noise occupation represents the number of thermal noise photons that would produce the same current fluctuations as the optically generated shot-noise.

The double-sided shot noise current spectral density is, 
\begin{equation}
S^{\mathrm{shot-noise}}_{II}(\Omega) = e_0\mathcal{R}{P}_0.
\end{equation}
We convert this to an effective photon noise occupation by comparing it to the quantum current noise spectral density of a thermal source~\cite{clerk2010introduction},
\begin{equation}
S^{\mathrm{thermal,q}}_{II}(\Omega) = \hbar\Omega \left(1 + 2\langle n \rangle\right) \frac{1}{Z_0},
\end{equation}
where $\langle n \rangle$ is given by the Bose-Einstein distribution. In this expression, $2\langle n \rangle$ represents the thermal fluctuations due to operation at a temperature above absolute zero. The broad shot-noise background is then indistinguishable (over a small fractional bandwidth around frequency $\Omega$) from operation at an elevated temperature with an effective occupation number,
\begin{equation}\label{eq:nth}
n_\text{eff}(\Omega) = \frac{1}{2}\left(\frac{e_0\mathcal{R} {P}_0 Z_0}{\hbar\Omega}\right).
\end{equation}
As expected, increased optical power, responsivity, and circuit impedance all lead to higher effective occupation number and noise temperature. Higher frequencies tends to reduce the effective occupation number since the occupation number for a fixed noise temperature goes with increasing frequency.

\subsection{Quantum Noise Limit of Qubit Gate Fidelity Due to Shot-Noise} 
\label{sec:gatefidelity}

Noise in the optical field affects not only the environment ($n_\mathrm{eff}$) but also qubit control.  Consider an ideal two-level system with transitions driven by an electric field or voltage $V(t)$. We assume this voltage is generated from a detected optical signal with power $P_\text{opt}(t)$ modulated at the qubit's transition frequency. The voltage is proportional to the optical power, $V(t) \propto P_0 f(t),$ where $f(t)$ is the pulse shape for performing the gate. In a simplified model, the Rabi frequency is proportional to $V(t)$, making the  qubit rotation given by the pulse area, $A$, proportional to the integrated optical power and therefore the total number of optical photons in the pulse $N_\text{opt,tot}$. Assuming that we want to perform an $X$-gate using a coherent laser with no other technical noise present, there will be a relative uncertainty in this area proportional to $(N_\text{opt,tot})^{-1/2}$, leading to a gate error (see~\hyperef{Appendix}{appendix:qubitgateerr}),
\begin{equation}
    \varepsilon_\text{QNL}=\left(\frac \pi 2 \frac{\delta A}{A}\right)^2=\left(\frac \pi 2\right)^2 \frac{1}{{N_\text{opt,tot}}}.\label{eqn:eps_QNL}
\end{equation}

\subsection{Efficiency-Noise Trade Off} 
\label{sec:tradeoff}
\hyperef{Sections}{subsec:figuresofmerit} and~\hyperef{}{sec:gatefidelity} reveal that we face a fundamental trade-off between power efficiency and noise: Improving the power conversion efficiency $\eta_\mathrm{P}$ reduces the number of optical photons $N_\text{opt,tot}$ required to perform a given gate, and therefore increases the gate infidelity caused by photon number fluctuations.

\begin{figure}[]
    \centering
    \includegraphics[width=\linewidth]{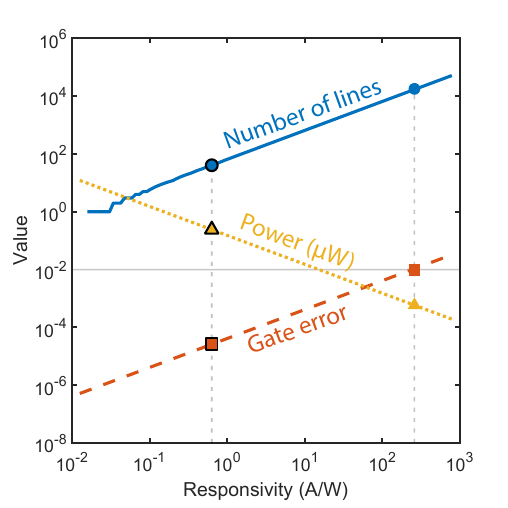}
    \caption{This plot shows the efficiency-noise trade-off. The solid blue line is the maximum number of qubits (\hyperef{Eq.}{eqn:Nqubit}), the red long-dashed line is the quantum noise limited $X$-gate error (\hyperef{Eq.}{eqn:eps_QNL}), and the yellow short-dashed line is the corresponding optical power required to perform the $X$-gate (\hyperef{Eq.}{eqn:poptpi}) in units of microwatts. To create this plot, we use $P_\text{cool} = \SI{10}{\micro\watt}$, $\tilde \Omega_\text{R}/(2\pi) = \SI{1.303}{\tera\hertz\per\volt}$, $\Delta t_\text{gate} = \SI{50}{\nano\second}$, and $Z_0 = \SI{50}{\ohm}$. The outlined markers are the operating parameters for a typical C-band photodiode responsivity $\mathcal{R} = \SI{0.84}{\A\per\watt}$: $P_0^{X} = \SI{180}{\nano\watt}$, $\varepsilon_\text{QNL} = 3.5\cdot 10^{-5}$, and $N_\text{qubit} = 50$. The unoutlined markers are for a system with increased responsivity $\mathcal{R} = \SI{240}{\ampere\per\watt}$: $P_0^{X}  = \SI{0.65}{\nano\watt}$, $\varepsilon_\text{QNL} =9.7\cdot10^{-3} \lesssim 10^{-2}$, and $N_\text{qubit} = 15000$.}
    \label{fig:scaling}
\end{figure}

\subsection{Strategies for Improvement}
\label{sec:improvements}
\subsubsection{Improving efficiency, responsivity, and impedance}
Our first challenge is improving efficiency so that more qubits can be optically addressed given a limited heat budget. Considering~\hyperef{Equation}{eqn:eta_p}, a clear strategy is to either improve the photodiodes by achieving higher quantum efficiency ($\eta_\mathrm{q}$), higher responsivity via external or internal amplification processes ($G$), or to increase the impedance of transmission line ($Z_0$) connected to the photodiode, \textit{e.g.}, by integrating high-impedance resonators or transformers with the photodiodes. These approaches have similar effects, so we only consider increases in total responsivity ($\mathcal{R})$.

Assume for a moment that novel detector technologies with improved responsivity are possible and can be implemented in a scalable way. In~\hyperef{Figure}{fig:scaling}, we plot the number of qubits (\hyperef{Eq.}{eqn:Nqubit}, blue solid line), gate error (\hyperef{Eq.}{eqn:eps_QNL}, red long-dashed line), and the optical power to perform an $X$-gate (yellow short-dashed line), as the photodiode responsivity is increased. Note that the optical power needed to perform an $X$-gate is computed by, 
\begin{equation}
    \label{eqn:poptpi}
    P_0^{X} = \frac{V_\pi}{\mathcal{R}Z_0} = \frac{\pi/(\tilde{\Omega}_\text{R}\Delta t_\text{gate})}{\mathcal{R}Z_0},
\end{equation} 
where $\tilde{\Omega}_\text{R}$ is the qubit Rabi frequency per volt, $\Delta t_\text{gate}$ is the gate duration, and $\mathcal{R} = \eta_\text{q}G\mathcal{R}_\text{q}$ is the total responsivity (see~\hyperef{Appendix}{appendix:qubitgateerr}). 

The plot shows a trade-off between the required optical power to perform the gate and the gate error, reminiscent of the shot-noise limit in optical measurements. If we hold gate error constant at $10^{-2}$, we can increase the number of qubits $N_\text{qubit}$ by three orders of magnitude from $10^1$ to $10^4$ while keeping the optical power around $10^{-10}$ watts, assuming that the approaches for improving the photodiode responsivity do not affect any other aspect of the system, \textit{e.g.}, lead to excess noise or power consumption. 

For reference, pulses used to perform single-qubit $X$-gates on superconducting qubits require average microwave powers ranging between $10^{-11}$ and $10^{-12}$ watts at the device for durations on the order of $50\cdot10^{-9}$ seconds~\cite{krinner2019engineering}. 

\subsubsection{Reducing fluctuations by nonlinear detection or intensity squeezed light}

Other strategies are needed to increase the number of drives while keeping the gate errors low. One approach is to use a nonlinear detector, \textit{e.g.}, by operating beyond the saturation power of the detector. In this regime, fluctuations of photon number around a certain value have a smaller effect. Alternatively, intensity-squeezed light could be used as a driving field while using standard square-law response detectors. Here, $\varepsilon_\text{QNL} \propto S^{-1}$, where $S$ is the amount of squeezing.

\subsubsection{Generating driving fields by coherent optical-to-microwave conversion}

An alternate approach, which in principle has no shot-noise, is to use a quantum transducer, such as an optomechanical transducer, to convert an incoming optical field coherently into a microwave field via a beam-splitter interaction. Such converters are under development to realize quantum connectivity between superconducting quantum machines~\cite{meesala2024non,jiang2023optically,delaney2022superconducting,han2021microwave,lambert2020coherent}. Quantum efficiencies on the order of unity have been demonstrated in several platforms~\cite{brubaker2022optomechanical, tu2022high, sahu2022quantum}.  A unity conversion efficiency would correspond to a power efficiency on the order of $\Omega/\omega_\mathrm{opt}\approx 10^{-5}$ for microwave systems. However, this efficiency only takes into account the power in the sideband. The process would need to be mediated by a much stronger optical pump field, likely $10^3$ larger than the sideband. This means that the overall power conversion efficiency would be at best $10^{-8}$ in this approach. Average microwave powers of $10^{-11}$ watts are typical for gate lines, which would correspond to an incoming optical power of $10^{-3}$ watts. To realize the targeted 10 microwatt of heating which enables a single gate line, all optical losses including absorption, scattering, and fiber-to-chip losses would need to be below $10^{-2}$, which is beyond the reach of current photonic technology.

\subsubsection{Prospects for increased RF frequency}
Now we consider a different perspective on optically-driven electrical signals at cryogenic temperatures. The expression for the power efficiency can be written in terms of the added noise from shot-noise,
\begin{equation}
    \eta_\text{P} = \eta_\text{q} \epsilon_\text{m}^2 \left(\frac{\Omega}{\omega_\text{opt}}\right)n_\text{eff}.
\end{equation}
In other words, if we increase $\Omega$ from 5 GHz (microwaves) to 100 GHz (mm-waves), for a fixed $n_\mathrm{eff}$, our power efficiency increases by a factor of 20. At the same time, the comparative advantage of optically driving mm-wave systems are accentuated by two additional factors:
\begin{enumerate}
    \item The cost and complexity of cryogenic mm-wave wiring and related components are significantly higher than that for microwave systems, so the simplicity of routing and multiplexing signals on optical fibers is attractive for scalability.
    \item Superconducting mm-wave systems and quantum devices can operate at higher temperatures, \textit{i.e.} 300 mK vs. 10 mK~\cite{anferov2024improved,anferov2024superconducting}. In a dilution refrigerator, the cooling power $P_\mathrm{cool}$ is proportional to $T^2_\mathrm{stg}$~\cite{wikus2010theoretical}, where $T_\mathrm{stg}$ is the stage temperature. This scaling means that much larger cooling power is available $P_\mathrm{cool}\approx 10^{-3}$ watts, allowing for an increased number of control lines \textit{i.e.} $N_\mathrm{qubit} \propto T_\mathrm{stg}^2$.
\end{enumerate}
Together, these three factors: lower noise, high mm-wave wiring cost and complexity, and the ability to operate at higher temperatures with greater cooling power, make optical driving at higher RF frequencies an attractive prospect in terms of its future scaling potential.  While the efficiency-noise trade off still exists at higher frequencies, it is less pronounced due to the increased cooling power. So, in this work we focus our effort in building and characterizing an optically-driven millimeter-wave source inside of a dilution refrigerator. In the following sections, we provide details on the cryogenic photogeneration of mm-wave signals and using these signals to probe a superconducting mm-wave resonator.

\section{\label{sec:srcdesc}Optically-driven Millimeter-wave Source}

\subsection{\label{subsec:princop} Principle of Operation}
Our goal is to build a mm-wave source that generates RF power by mixing two coherent optical fields with different frequencies on a high-speed photodiode (HSPD) in a 4 K environment. 

To characterize the source and measure superconducting mm-wave circuits we attach an adapter that converts the RF signals from the coaxial output of the HSPD into a rectangular waveguide. We do this because rectangular waveguides have lower loss than coaxial cables at mm-wave frequencies~\cite{stokowski2019towards,multani2020development,anferov2020millimeter}.~\hyperef{Figure}{fig:princop} illustrates the spectrum at the HSPD input and waveguide output, emphasizing that the output spectral bandwidth set by the cutoff frequency of the rectangular waveguide and and bandwidth of the HSPD.

\begin{figure}[!ht]
\centering
\includegraphics[scale=1.0]{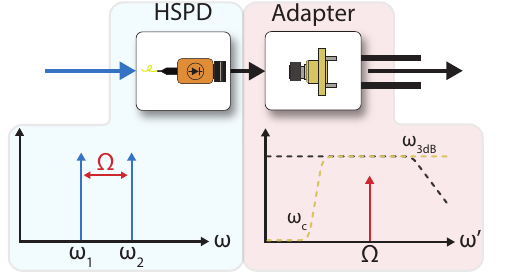}
\caption{\label{fig:princop} Schematic of our optically-driven source. The blue arrow represents the optical path, and the black arrow represents a coaxial mm-wave path. The double-black line indicates a waveguide mm-wave path. The bottom left graph (blue shading) shows the power spectral density of two optical tones impinging onto the high-speed photodiode (HSPD). The bottom right graph (red shading) shows the power spectral density at the output of the coax-to-waveguide adapter. The black dashed line is the response of the HSPD, and the gold dashed line shows the waveguide response.}
\end{figure}

\subsection{Cryogenic Power and Noise Characterization}
\label{subsec:charac}
By mounting the HSPD, a W-band harmonic mixer, and a cryogenically calibrated zero-bias GaAs mm-wave detector in a dilution refrigerator (Bluefors, LD-250), we characterized the signal power and broadband noise power generated by the source.~\hyperef{Figure}{fig:pwrnoise} (c) shows a schematic of our setup, which we used to characterize both quantities. To generate the mm-waves, we modulated a laser with a QPSK electro-optic modulator (Fujitsu FTM7961EX) driven by a signal generator (SG1, Keysight E8257D). We null-biased the EOM with a bias controller (BC, Plugtech MBC-IQ-03) to maximize power in the first-order sidebands. From the output of the EOM, the modulated light was amplified by an Erbium-doped fiber amplifier (EDFA) and sent into the dilution fridge via optical fiber to the HSPD on the 4 K stage. From here, the mm-wave signal is generated and sent to one of two paths used in separate cooldowns, as shown in~\hyperef{Figure}{fig:pwrnoise} (c). We used the the first path, labeled (1) to measure the generated mm-wave power, and the second path, labeled (2), to measure the noise power spectral density and mm-wave superconducting circuits. 
\begin{figure}[ht!]
\centering
\includegraphics[]{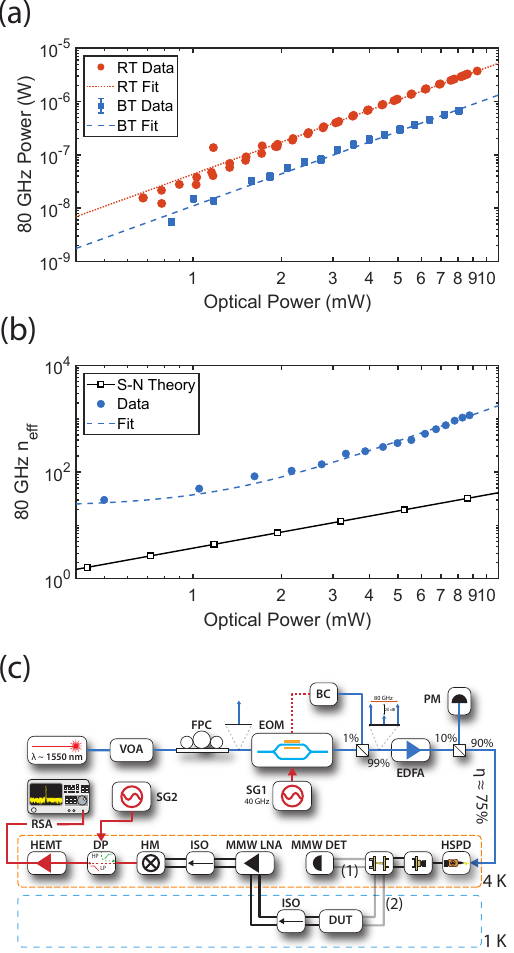}
\caption{\label{fig:pwrnoise} (a) 80 GHz power at $\SI{298}{\K}$ (red) and at $\SI{3.09}{\K}$ (blue) measured using path (1). The solid points indicate data, while the dashed lines are quadratic fits. Error bars indicate standard deviation. (b) 80 GHz added thermal photons as a function of optical power at $\SI{3.31}{\K}$ measured using path (2) in a separate cooldown. The solid black line with square markers shows the expected noise level, assuming we are limited by shot noise.  It is plotted using~\hyperef{Equation}{eq:nth}, given $\mathcal{R} = \SI{0.05}{\ampere\per\watt}$. (c) Schematic of the measurement set-up. The blue lines indicate the optical signal path, the black single lines denote the mm-wave signal path connected by 1.85 mm coaxial cables, the double black lines show mm-wave paths connected by rectangular waveguides, and the red lines indicate the microwave signal path.}
\end{figure}

To characterize the mm-wave power generated by the source, we drove the null-biased EOM at 40 GHz so that the first-order sidebands would generate an 80 GHz signal. After the sidebands mixed on the HSPD, the mm-wave signal continued through a 10-inch long stainless steel 1.85 mm coaxial cable (RF Coax, V086MMSB-10R) to a 1.85 mm-to-WR15 adapter (Eravant), then through a WR15-WR10 waveguide adapter (Eravant). After the WR10 adapter, the signal is detected by a zero-bias GaAs W-band detector (MMW DET, Pacific Millimeter Products WDH-SP), which we calibrated at cryogenic temperatures (see~\hyperef{Appendix}{appendix:wbandcal}). Using this setup, we measured the mm-wave power as a function of optical power, shown in~\hyperef{Figure}{fig:pwrnoise} (a). We performed the optical power sweep at room temperature (red data) and cryogenic temperatures (blue data) using a voltage-controlled optical attenuator (VOA). During these measurements, we observed no appreciable heating of the dilution refrigerator stages (see~\hyperef{Appendix}{appendix:bluefors}).

Since we measured the mm-wave power after the WR15-WR10 tapered transition, the insertion loss of the components before are included in our observations. The total insertion loss from the output of the HSPD to the input of the detector by,
\begin{equation}
    |\mathcal{L}(\omega, T)|^2 = \frac{a(\omega, T)}{\mathcal{R}^2(T)\cdot Z_0},
\end{equation}
where $\mathcal{L}(\omega, T)$ is the total insertion loss as a function of RF frequency and temperature, and $a$ is a fit parameter to the function $f(x) = 1/2ax^2$ used to fit the power data in~\hyperef{Figure}{fig:pwrnoise} (a). This equation compares the coefficients of the measured data to the expected quadratic relationship discussed in~\hyperef{Section}{subsec:figuresofmerit} and ~\hyperef{Section}{subsec:princop}. If there were no insertion loss between the output of the HSPD and the mm-wave detector, then $\mathcal{L}(\omega, T) = \mathcal{H}(\omega, T)$, the HSPD transfer function. 
\begin{table}[h]
\caption{Insertion loss estimates from HSPD output to mm-wave detector.}
\label{tab:insertionloss}
\begin{ruledtabular}
\begin{tabular}{cccc}
$T \text{ (K)}$ &$\mathcal{R} \text{ (AW}^{-1})$ &  $a \text{ (W}^{-2})$  & $|\mathcal{L}| \text{ (dB)}$ \\
\hline
298  & $0.40 \pm 0.02$ & $0.090 \pm 0.001$ & $-20 \pm 0.4$ \\
3.09 & $0.050 \pm 0.003 $ & $0.020 \pm 0.0003$ & $-8 \pm 0.4$ \\
\end{tabular}
\end{ruledtabular}
\end{table}

\hyperef{Table}{tab:insertionloss} shows the measured HSPD responsivity, the fit parameter $a$, and the total insertion loss estimates for two different temperatures. We find that the responsivity changes as a function of temperature, which we attribute to thermal effects introducing misalignment to the optical-fiber-photodiode interface~\cite{bardalen2018evaluation}. The $12$ dB difference in insertion loss is attributed to the increased electrical conductivity of the RF cables and waveguides at $\SI{3.09}{\K}$. 

We find minimal change in the HSPD bandwidth~\cite{Zhang1997cryogenic, bardalen2018evaluation} from room temperature down to cryogenic temperatures in these measurements. We provide further details on our room temperature HSPD characterization in~\hyperef{Appendix}{appendix:hspdcharacterization}.

In a separate measurement, we investigate the broadband noise generated by the source using path (2). After the WR15 to WR10 transition, the signal was routed through a standard WR10 waveguide section (DUT). Afterwards, we send the signal into the detection chain. The detection chain consists of a mm-wave low noise amplifier (MMW LNA, Low Noise Factory LNC65\_115WB) with cryogenic isolators (ISO, MicroHarmonics FR100C) before and after the amplifier. The waveguide section that joins the first isolator and the MMW LNA is stainless steel, to avoid a thermal short between the 1 K and 4 K stage. After the second isolator, we connect a harmonic mixer (HM, MI-WAVE 920W/387) to a diplexer (DP, MiniCircuits ZDSS-5G6G-S+). To measure the 80 GHz signal we mix it down using the HM to an IF frequency of 500 MHz and measure the power spectral density on a real-time spectrum analyzer (RSA) as a function of optical power. We drive the harmonic mixer at $\fLO = \frac{1}{N}\left(f_\text{RF}-f_\text{IF}\right) = 5.6786$ GHz, where $N=14$ is the harmonic factor for the mixer (SG2, Keysight E8257D). From these measurements we can calibrate the total power gain of the detection chain, $G_\text{tot}=0.53$ and measure the single-sided 500 MHz noise power spectral density (PSD), $S^\text{ss, IF}$ in watts per hertz. The total detection gain relates the measured IF power to the generated mm-wave power, $P_\text{IF} = G_\text{tot}P_\text{RF}$.

The solid blue circles in \hyperef{Figure}{fig:pwrnoise} (b) denote the effective thermal photon number at 80 GHz as a function of optical power. We calculate this by converting the measured single-sided 500 MHz noise PSD to a double-sided 80 GHz noise PSD, $S^\text{ds, RF}_{II} = S^\text{ss, IF}/(2G_\text{tot}Z_0)$, with $Z_0 = 50 \text{ }\Omega$. To convert this to $n_\text{eff}$ we use, $n_\text{eff} = \frac{1}{2}(S^\text{ds, RF}_{II}Z_0)/({\hbar\Omega})$, where $\Omega/(2\pi)=80\text{ GHz}$. The measured noise scales nonlinearly with optical power, which we fit with $f(x) = ax^2+b$, where $a = 1.5\cdot10^{7} \text{ photons/W}^2$ and $b = 23 \text{ photons} $. This suggests the presence of excess noise in our measurements~\cite{putra2017edfanoise, lecocq2021control}. Moreover, we plot the expected noise level in \hyperef{Figure}{fig:pwrnoise} (b), assuming we are limited only by shot noise, with the solid black and square markers. We observe that the measured noise is about an order of magnitude higher than would be expected in a quantum-limited setting. For the theory plot we have assumed a responsivity of $\mathcal{R} = \SI{0.05}{\ampere\per\watt}$ which we measured at 3.09 K. The minimum and maximum measured 80 GHz noise power is equivalent to $n_\text{eff}^{\text{min}}= 30$ for \SI{0.5}{\milli\watt} of optical power and $n_\text{eff}^{\text{max}}= 1200$ for \SI{8.7}{\milli\watt} of optical power (see~\hyperef{Appendix}{appendix:mixernoise}.)

\section{Superconducting Millimeter-wave Circuits}
\label{sec:scmmw}
We fabricated superconducting mm-wave circuits and characterized them using the source. These circuits are quarter-wave resonators made from 100 nm thick NbTiN, sputtered onto a $\SI{195}{\micro\meter}$ thick sapphire substrate (STAR Cryogenics). A microscope image of the resonators is shown in~\hyperef{Figure}{fig:scmmw} (a). We pattern the device using photolithography and etch using a Chlorine-based reactive ion etch.~\hyperef{Figure}{fig:scmmw} (b) and (c) show the final dimensions of the device chip and how it is packaged into a WR10 rectangular waveguide, respectively. The copper holder secures the chip so that the normal vector of the NbTiN plane is parallel to the broad wall of the rectangular waveguide. This ensures that the fundamental TE mode of the waveguide can sufficiently couple to the dipole moment of the superconducting electrodes, driving the resonators. In these measurements, we used path (2) shown in~\hyperef{Figure}{fig:pwrnoise} (c), with the DUT now being the packaged chip.

Because the maximum output frequency of our signal generator was 40 GHz (SG1), we used the second-order sidebands of the modulated light to generate the frequencies necessary to characterize the cavities. In this scheme, to cover the W-band (75 GHz -- 110 GHz), the modulation frequencies must be between 18.75 GHz and 27.50 GHz. So, if we drive a null-biased EOM at $\Omega$, we generate sidebands $\omega_c \pm \Omega$ and $\omega_c \pm 2\Omega$, where $\omega_c$ is the laser carrier frequency. This generates RF beat tones at the output of the HSPD at $\Omega, 2\Omega, 3\Omega, 4\Omega$, where the desired signal is $f_\text{RF} = 4\Omega$. Our system's highest cutoff frequency is due to the WR10 waveguide at 59 GHz. For the frequencies of interest, $\Omega$ and $2\Omega$ are always below the cutoff, so only $3\Omega$ and $4\Omega$ propagate.

After the desired mm-wave frequency is generated, it is guided to the device and down-converted to a fixed intermediate frequency $f_\text{IF}$, using the harmonic mixer, by driving it at a local oscillator (LO) frequency, $f_\text{LO}$, as explained in~\hyperef{Section}{subsec:charac}. Because the IF frequency is fixed, the relationship between the EO modulation frequency and the LO frequency is $f_\text{IF} = f_\text{RF} - N f_\text{LO} = 4\Omega - 8 f_\text{LO}$. To perform mm-wave spectroscopy, we adjust $\Omega$ and $f_\text{LO}$ accordingly (SG1 and SG2 frequencies, respectively). Note that it is possible to do this process with only one signal generator instead of two, by building an additional RF amplification chain (see~\hyperef{Appendix}{appendix:slfi}). In selecting the harmonic mixer drive frequency in this way, we ensure that the measured power spectral density has no component from the $3\Omega$ leakage.

\hyperef{Figure}{fig:scmmw} (d) shows the result of the frequency domain spectroscopy. We expect an asymmetric Lorentzian line shape in these measurements due to mismatched input and output impedances, as seen from the resonator~\cite{anferov2020millimeter}. We measure four different cavities with different electrode lengths to vary the resonance frequency.~\hyperef{Figure}{fig:scmmw} (e) shows a zoomed-in portion of the lowest frequency cavity and a corresponding Lorentzian fit. From the fit, we can deduce the external quality factor to be $Q_{\text{e}} = 2 \cdot 10^3$ and the internal quality factor to be $Q_\text{i} = 6 \cdot 10^4$.

\begin{figure}[h!]
\centering
\includegraphics[]{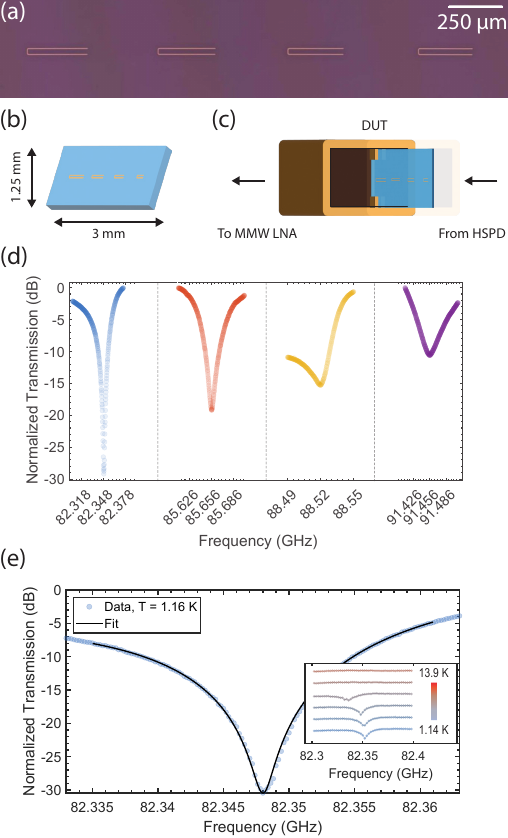}
\caption{\label{fig:scmmw} (a) Optical microscope image of the 100 nm thick NbTiN cavities. The gap between the electrodes is 22 $\mu$m, and the lengths of the electrodes from left to right are 275 $\mu$m, 265 $\mu$m, 255 $\mu$m, 245 $\mu$m. (b) A graphic indicating the in-plane dimensions of the double-side polished sapphire substrate on which the cavities are fabricated. (c) A graphic of the packaging of the full device. The sapphire die is glued inside of a custom holder, establishing a fin-line geometry. (d) Frequency domain spectroscopy of all the cavities using the optically-driven source. (e) Zoom in on the lowest frequency mode, where $\omega_0/2\pi\approx$ 82.348 GHz and $Q\approx 2\cdot 10^3$. The inset shows measurements taken while warming up the cryostat. We observe a clear signature of a superconducting-to-normal phase transition.}
\end{figure}

\section{Conclusion}
\label{sec:conclusion}
Controlling large numbers of superconducting qubits is rapidly becoming one of the major challenges in scaling up nascent quantum computers to the scales required for fault-tolerant algorithms. Optical interconnects have been proposed as a solution~\cite{lecocq2021control,li2024optical}, yet we find that these approaches present an efficiency-noise trade-off that make them challenge to use at millikelvin temperatures. Moving to higher temperatures and higher frequencies may alleviate some of these constraints. As a proof of principle, we demonstrate and characterize a cryogenic optically-driven mm-wave source. We used this source to perform frequency domain spectroscopy of superconducting resonators in the W-band (75 - 110 GHz). Additionally, we measured the mm-wave power generated and broadband noise characteristics of the source as a function of optical power in a 4 K environment. Operating the source with roughly 10 mW of modulated optical power modulated, we were able to generate $\sim\SI{1}{\micro\watt}$ at 80 GHz with about $1500$ photons of added noise. We also observed that the measured noise had a nonlinear dependence with increasing optical power, indicating the presence of excess noise, which we expect can be reduced by using a lower noise optical driving circuit. 

Besides qubit control, these sources may be more suitable in other applications of quantum technologies, such as serving as pumps to drive parametric processes in superconducting mm-wave circuits~\cite{stokowski2019towards,anferov2020millimeter}, for frequency conversion~\cite{pechal2017millimeter, kumar2023quantum}, amplification~\cite{tan2024operation}, and quantum sensing~\cite{vaartjes2023strong}.

\begin{acknowledgments}
K.K.S.M. and W.J. would like to acknowledge Yudan Guo and Nathan Lee for helpful discussions. A.S.-N. acknowledges useful conversations with Florent Lecocq, John Teufel, Scott Diddams, Frank Quinlan, Mohammad Mirhosseini, and Oskar Painter. K.K.S.M. gratefully acknowledges support from the Natural Sciences and Engineering Research Council of Canada (NSERC). Part of this work was performed at the Stanford Nano Shared Facilities (SNSF) and Stanford Nanofabrication Facility (SNF), supported by the National Science Foundation under award ECCS-2026822. Supported in part by the Department of Energy Contract No. DEAC02-76SF00515.
\end{acknowledgments}

\appendix
\section{Derivation of Qubit Gate Error Due to Shot Noise}
\label{appendix:qubitgateerr}
In this section we sketch the relationship between qubit gate error and shot noise, as discussed in Section~\ref{sec:gatefidelity}. For a general rotation about the $x$-axis of the Bloch sphere, we can write
\begin{equation}
\hat{R}_{X}(\theta) = \cos(\theta/2)\hat{\mathbf{1}} - i\sin(\theta/2)\hat{\sigma}_x,
\end{equation}
where $\hat{\mathbf{1}}$ is the identity operator and $\hat{\sigma}_x$ is the Pauli X operator.

We are interested in what an imperfect $\pi$-pulse would do to the ground state $\ket{0}$. If we let $\theta = \pi + \delta\theta$ with $\delta\theta/\pi \ll 1$ and neglect $\mathcal{O}(\delta\theta^2)$ terms, we obtain the relation
\begin{equation}
    \hat{R}_{X}(\pi+\delta\theta)\ket{0} \approx -\frac{1}{2}\delta\theta \ket{0} - i \ket{1}.
\end{equation}
So the error probability is, 
\begin{equation}
\label{eq:error}
    P_\text{error} = \left(\frac{\delta\theta}{2}\right)^2.
\end{equation}
This error probability is the probability that we measure $\ket{0}$ given that we prepared $\ket{1}$ and is equal to the probability that we measure $\ket{1}$ given that we prepared $\ket{0}$.

Now consider, the two-level system driven by an electric field or voltage $V(t)$, like in the case of a transmon qubit coupled to an XY drive line. By selecting the microwave drive frequency and phase carefully, we can coherently control the qubit so that we can generate rotations about the $x$-axis of angle $\theta$~\cite{blais2004circuitqed,krantz2019quantumeng}. In this case, we can write the rotation angle as
\begin{equation}
    \theta(t) = \tilde{\Omega} V_0 \int_{0}^{t} f(t') \mathrm d t',
\end{equation}
where $\tilde{\Omega}_\text{R} = (C_{\text{xy}}/C_\Sigma) Q_\text{zp}$ denotes the Rabi frequency per volt and $V(t) = V_0 f(t)$, where $V_0$ is the drive voltage and $f(t)$ is a dimensionless envelope function. Here $C_\text{xy}$ and $C_{\Sigma}$ are the capacitance of the qubit to the microwave drive line and total capacitance, respectively. Also, $Q_\text{zp} = \sqrt{\hbar/(2Z_\text{q})}$, where $Z_\text{q} = 1/(\omega_\text{qubit}C_\Sigma)$ is the impedance of the transmon. 

Now, suppose the pulse is generated by modulating coherent optical light impinging on a photodiode, mounted in the dilution refrigerator. Then, we can write $V_0 = Z_0 I_\text{PD} = Z_0 \mathcal{R}P_0 = Z_0 G\eta_\text{q}\mathcal{R}_\text{q}P_0 = Z_0 G\eta_\text{q} \mathcal{R}_\text{q} \hbar\omega_\text{opt} N_\text{tot,opt}/\Delta t_\text{gate}$, where $Z_0$ is the characteristic impedance of the environment of which the photodiode is coupled and $N_\text{tot,opt}$ is the total number of optical photons in the pulse. Here we assume the envelope function is rectangular with unit height and width $\Delta t_\text{gate}$. The gate angle can be written as,
\begin{equation}
    \theta = \tilde{\Omega}_\text{R}Z_0\mathcal{R}\hbar\omega_{\text{opt}} \cdot N_\text{tot,opt}.
\end{equation}
Because we are utilizing a coherent optical drive we know that $\delta N_\text{tot,opt} = \sqrt{N_\text{tot,opt}}$
\begin{equation}
    \delta\theta = \left(\tilde{\Omega}_\text{R}Z_0\mathcal{R}\hbar\omega_{\text{opt}}\right)\cdot \sqrt{N_\text{tot,opt}}.
\end{equation}
In the case of a $\pi$-pulse, the target angle is $\theta = \pi$, implying $\pi/N_\text{tot,opt} = \tilde{\Omega}_\text{R}Z_0\mathcal{R}\hbar\omega_{\text{opt}}$. So we have,
\begin{equation}
    \delta\theta = \frac{\pi}{\sqrt{N_\text{tot,opt}}}
\end{equation}
Therefore using \hyperef{Equation}{eq:error} the probability of an error occurring after a $\pi$-pulse using optically driven techniques is,
\begin{equation}
    P_\text{error} = \left(\frac{\pi}{2}\right)^2\frac{1}{N_\text{tot,opt}}.
\end{equation}
To plot \hyperef{Figure}{fig:scaling} we used $C_\text{xy}=\SI{0.3}{\femto\farad}$, $C_\Sigma = \SI{200}{\femto\farad}$, $\omega_\text{qubit}/(2\pi) = \SI{5}{\GHz}$, and $\Delta t_\text{gate} = \SI{50}{\nano\second}$. These numbers give $\tilde\Omega_\text{R}/(2\pi) = \SI{1.303}{\tera\hertz\per\volt}$.

\section{Cryogenic Calibration of W-band Detector}
\label{appendix:wbandcal}
In order to measure the power output of our optically-driven source, benchmarked a mm-wave diode power detector at cryogenic temperatures. We purchased a GaAs zero-bias diode detector from Pacific Millimeter-wave Products and characterized its sensitivity at 298 K and at 5 K. The sensitivity at cryogenic temperatures was characterized in three steps: (1) Assume the room temperature sensitivity of the detector and measure the power of our mm-wave frequency extender (2) Measure the insertion loss of the test setup at 298 K and at 5 K (3) Put the detector at a known location in the setup and measure the sensitivity at 5 K.

The detector's datasheet tells us the sensitivity as a function of frequency and the square-law saturation power. At 298 K, we measured the power of our mm-wave frequency extender as a function of frequency. Additionally, we varied the attenuation level going to the detector using a variable WR10 attenuator at each frequency. \hyperef{Figure}{fig:mmwdet} (a) shows the result of this measurement. The light blue data shows the power at the extender output port after removing the effect of the attenuation. The black line shows a typical frequency extender response.

\begin{figure}[!htbp]
\centering
\includegraphics[scale=1.0]{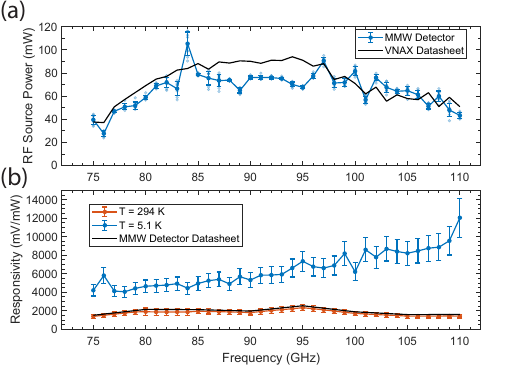}
\caption{\label{fig:mmwdet} (a) Output power of the VDI WR15 VNAX as a function of frequency, measured by the mm-wave detector from Pacific Millimeter Products (PMP). The blue points show the measured power, while the black line shows a typical output power spectrum from a generic data sheet (VDI). (b) Sensitivity of the zero-bias GaAs diode detector at $T = \SI{294}{\K}$ (red) and $T= \SI{5.1}{\K}$ (blue). The black line shows the datasheet for the detector's sensitivity from PMP.}
\end{figure}

We built a symmetric rectangular waveguide setup so that we can measure the insertion loss as a function of temperature. By ``symmetric", we mean that we use the same components at the input and output of the measurement chain. Then, at the symmetric point,  we assume half of the measured insertion loss has been accumulated. We built this setup in a Montana Instruments cryostat. After characterizing the insertion loss, we install the mm-wave detector at the symmetric point. Then, we perform the power and frequency measurements as before. \hyperef{Figure}{fig:mmwdet} (b) shows the result of these measurements. The blue data is the sensitivity measured at base temperature. The red data shows the measured sensitivity at room temperature, and the light grey line is the data sheet sensitivity. 

In the main text, we use the sensitivity at 80 GHz, $\mathcal{S}=4649 \pm \SI{673}{\milli\volt\per\milli\watt} $ to deduce the output power generated by the optically-driven source. Note this method only works if we are beating the first-order sidebands because the mm-wave spectrum has no other frequency components. 
\section{Harmonic Mixer Gain and Noise Figure Estimates}
\label{appendix:mixernoise}
As depicted in~\hyperef{Figure}{fig:pwrnoise} (a) the detection chain includes a cryogenic low noise mm-wave amplifier (LNF-LNC65\_115WB), a harmonic mixer (MI-WAVE 920W/387), and a cryogenic microwave low noise amplifier (LNF-LNC0.2\_3B). The amplifiers' respective data sheet values are tabulated in~\hyperef{Table}{tab:amps}.

\begin{table}[h]
\caption{Cryogenic gain and noise figure for the mm-wave and microwave low noise amplifiers.}
\label{tab:amps}
\begin{ruledtabular}
\begin{tabular}{cccc}
P/N & $f \text{ (GHz)}$ &  $G \text{ (dB)}$  & $F \text{ (dB)}$ \\
\hline
\small{LNF-LNC0.2\_3B} & 0.5 & $31$ & 0.03 \\
\small{LNF-LNC65\_115WB} & 80 & $22.5$ & 0.3 \\
\end{tabular}
\end{ruledtabular}
\end{table}

We note that microwave power and the mm-wave power are directly proportional, $P_\text{500 MHz} = G_\text{tot} P_\text{80 GHz}$. We measure $G_\text{tot} = 0.53$ or $-2.76$ dB. Therefore, the harmonic figure's conversion gain is no more than $G_\text{tot}/(G_\text{mm-wave}G_{\mu\text{w}}) = 2.29 \cdot 10^{-6}$ or $-56.3$ dB.

Using the Friis formula for cascaded noise we write the total noise figure of the detection chain,
$$
    F_\text{tot} = F_\text{mmw} + \frac{F_\text{mixer}-1}{G_\text{mmw}} + \frac{F_\text{µw}-1}{G_\text{mmw}G_\text{mixer}}, 
$$
where $F$ denotes the noise factor of each respective component and $G$ is the gain. Noise factor is defined as the ratio of the input SNR and the output SNR, $F = \text{SNR}_\text{in}/\text{SNR}_\text{out}$. For us, the input of the detection chain is the output port of the HSPD and the output of the detection chain is the input of the RSA. The data sheet provides values for the gain and noise factor of each of the cryogenic low noise amplifiers, so we can solve for the mixer's noise factor,
$$
    F_\text{mixer} = 1 + G_\text{mmw}\left(F_\text{tot} - F_\text{mmw}\right) - \frac{F_\text{µw} - 1}{G_\text{mixer}}.
$$
To estimate the mixer's noise factor, we need to calculate the total noise figure of the detection chain \textit{i.e.} $F_\text{tot} = \text{SNR}_\text{in}/\text{SNR}_\text{out}$. In experiment, we determined $\text{SNR}_\text{out}$ by recording the noise and signal power on the RSA. For $P_0 \approx 1 \text{ mW}$, we obtain $\text{SNR}_\text{out}$ $12.6$ or $11.0 \text{ dB}$. 

The input SNR is determined by the input signal power and the total input noise power. We assume the input noise power is composed of $4 \text{ K}$ Johnson-Nyquist noise and shot noise. For $P_0 \approx 1 \text{ mW}$, the shot-noise power is $0.0167 \text{ nW}$, and the thermal noise power is $0.00116 \text{ nW}$, where we used the bandwidth $80.0 \text{ GHz} - 59.0 \text{ GHz} = 21.0 \text{ GHz} $, defined by the lower cut-off frequency of the WR10 rectangular waveguide and by the 3 dB bandwidth of the HSPD. The signal power is $P_\text{80 GHz} = 14.97 \text{ nW}$. Therefore, the input SNR is $\text{SNR}_\text{in} = 838$ or $29.2 $ dB, and the total noise factor is $F_\text{tot} = 76.2$ or $18.8 \text{ dB}$. Finally, we compute the mixer's noise factor using the Friis formula, as $\approx 41$ dB.

\section{Same LO, fixed IF (SLFI) Measurement}
\label{appendix:slfi}
In the main text, we explained the readout chain for the noise measurement and the superconducting cavity measurement. In those measurements, we used two different signal generators, SG1 and SG2 (see~\hyperef{Figure}{fig:HSPD}). One signal generator is to drive the electro-optic modulator, and the other is to drive the harmonic mixer. In this section, we describe how to use just one signal generator for both the mm-wave generation and readout; we are calling this SLFI: same LO, fixed IF, alluding to the fact that the same local oscillator (LO) is being used for the signal up-conversion and down-conversion so that the intermediate frequency (IF) is fixed. This can also be seen as a type of superheterodyne scheme.

We built an up-conversion setup to control the optical sidebands generated by the modulator. The local oscillator (LO) signal is generated by an analog signal generator (Keysight E8257D 250 kHz - 40 GHz) at frequency $f_\text{LO}\sim 10~\text{GHz}$. The LO is split to feed both an IQ mixer (Marki microwave IQ0618) and the harmonic mixer for down-conversion (MI-WAVE 920W/387). We use IF signals at $f_\text{IF}=$ 50 MHz from an arbitrary waveform generator (Rigol DG4102, 100 MHz) as the IQ input with optimized phase and amplitude such that the RF output of the IQ mixer is dominated by a single up-converted sideband. The RF signal at $f_\text{RF} = f_\text{LO} + f_\text{IF}$ is then amplified (Mini-Circuits, ZX60-06183LN+), frequency-doubled (Mini-Circuits, ZXF90-2-44-K+), and amplified (Mini-Circuits, ZX60-24-S+) again before the electro-optic modulator. The electro-optic modulator (Fujitsu, FTM7961EX) operates in the null-biased mode, where we are using the second-order sidebands to generate the mm-wave signal. This null-biased setting is controlled by a modulator bias controller (PlugTech MBC-IQ-03). A 99:1 beam splitter after the modulator provides feedback for the bias controller. The output of the beam splitter is amplified by an erbium-doped fiber amplifier (EDFA, FiberPrime EDFA-C-26G-S) before entering the dilution refrigerator.

\begin{figure}[!htbp]
\centering
\includegraphics[scale=1.0]{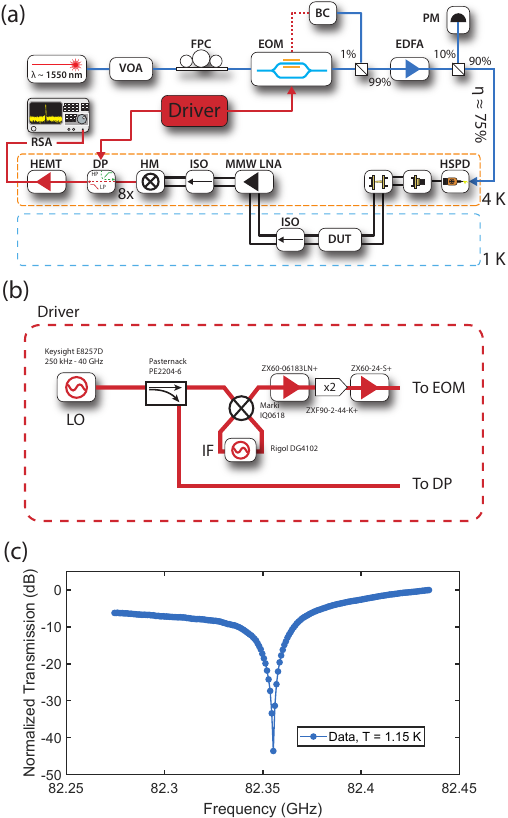}
\caption{\label{fig:slfi} (a) Schematic of the SLFI setup. Note most of the setup is similar to what is shown in~\hyperef{Figure}{fig:HSPD}, with the exception of the driver block. The purpose of the driver block is to use the same source to drive the EOM and the down-converting harmonic mixer. (b) A measurement of the same superconducting cavity is shown in \hyperef{Figure}{fig:scmmw} (e), except with the SLFI setup.}
\end{figure}

Because of the frequency limits of the microwave components, we used the second-order sidebands from the modulator to generate the mm-wave signal used for the device characterization. As a result, the mm-wave frequency $f_\text{mmw} = 8f_\text{RF}$. The output noise is sent through a circulator (Microharmonics, FR100C), a mm-wave cryogenic low noise amplifier (Low Noise Factory, LNF-LNC65\_115WB), and to the harmonic mixer. We use the same LO for the down-conversion at harmonic number 8. The resulting IF signal at the output of the harmonic mixer is $f_\text{o} = 8f_\text{IF}=400~\text{MHz}$. We record the output signal power using a spectrum analyzer (Rohde \& Schwarz FSW). As we sweep the LO frequency, the mm-wave frequency sweeps across the mm-wave resonators without affecting the output signal frequency.

\section{Dilution Fridge Stage Heating}
\label{appendix:bluefors}
In Sections II and III, we describe operating the source in a dilution refrigerator. Below, we show a plot of the dilution fridge (Bluefors LD-250) stage temperatures while we were conducting the base temperature power sweeps, shown in \hyperef{Figure}{fig:pwrnoise} (a) (blue data). We observe no stage heating correlated with our measurements. We do not observe fluctuations in the MXC temperature because the thermometer reading stayed below its 7 mK threshold. 

\begin{figure}[htpb]
\centering
\includegraphics[]{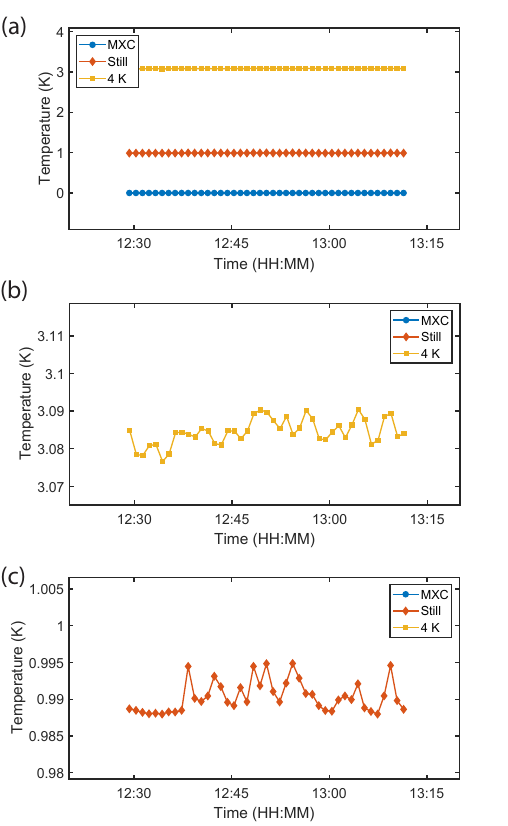}
\caption{\label{fig:bftemps} (a) Temperatures of the stages in the dilution refrigerator while conducting source power sweeps. The MXC value did not change throughout the measurement, indicating that the MXC temperature reading did not go above 7 mK. (b) Zoomed-in data of the 4 K stage. (c) Zoomed-in data of the Still stage. }
\end{figure}

\section{High-Speed Photodiode Characterization}
\label{appendix:hspdcharacterization}
As the HSPD (II-V XPDV2120R-VF-FA) is a critical component in our optically-driven mm-wave source, we built an experimental setup to benchmark its properties.~\hyperef{Figure}{fig:HSPD} (a) shows a schematic of the setup. At room temperature, two lasers were combined at a 50:50 beam splitter and then sent to the HSPD. The output of the HSPD was routed to one of two mm-wave receivers via a 1.85 mm coaxial cable. Subsequently, the receiver down-converted the mm-wave signal to approximately 3 GHz. We then recorded the power spectral density using a real-time spectrum analyzer (RSA, Rhode \& Schwarz) as a function of the laser detuning (which changes the RF frequency). The results of this measurement are shown in~\hyperef{Figure}{fig:HSPD} (c). We found the 3-dB bandwidth to be $f_\text{3dB} \approx$~75 GHz.

\hyperef{Figure}{fig:HSPD} (b) shows a diagram of the two mm-wave receivers we used. One uses a V-band frequency extension module (Virginia Diodes, WR15-VNAX), and the other uses a W-band harmonic mixer (MI-WAVE 920W/387). We built these two receivers because we were interested in comparing their performance at room temperature before mounting one of them in the dilution refrigerator.~\hyperef{Figure}{fig:HSPD} (c) shows that these two receivers agree relatively well at room temperature. However, the harmonic mixer's conversion loss varies with frequency, leading to inconsistent output power. Due to the harmonic mixer's lower power consumption and smaller size (no DC bias, 3 in x 3 in x 1 in), we selected it to be mounted in the dilution refrigerator for cryogenic measurements (\hyperef{Section}{subsec:charac},~\hyperef{Section}{sec:scmmw})~\footnote{In comparison, the frequency extender has a larger footprint and higher power demand (9 V/45 W, 3 in x 13 in x 3 in).}.

To investigate the HSPD's responsivity as a function of temperature, we mounted it inside a 4 K cryostat (Montana Instruments). To simplify the cryogenic setup, we measured beat frequencies in the microwave frequency range directly on the RSA to stay well within the photodiode's bandwidth, as indicated in~\hyperef{Figure}{fig:HSPD} (a). \hyperef{Figure}{fig:HSPD} (d) shows the measurement of DC and RF responsivity as a function of temperature. We performed an optical power sweep for each data point by varying a voltage-controlled optical attenuator (VOA). We recorded the DC photocurrent and RF power spectral density at each attenuation setting and temperature. The DC and RF responsivities show 10 dB extra insertion loss in a 4 K environment. Notably, this insertion loss is recoverable as the responsivity curves show no signs of hysteresis, denoted by the \textit{warmup} and \textit{cooldown} labels in~\hyperef{Figure}{fig:HSPD} (d). This indicates that there is no permanent damage due to the temperature cycling. We attribute the responsivity drop to extra insertion loss due to the thermal contraction and expansion effect at the optical-fiber-photodiode interface~\cite{bardalen2018evaluation}.

\begin{figure*}[!ht]
\centering
\includegraphics[]{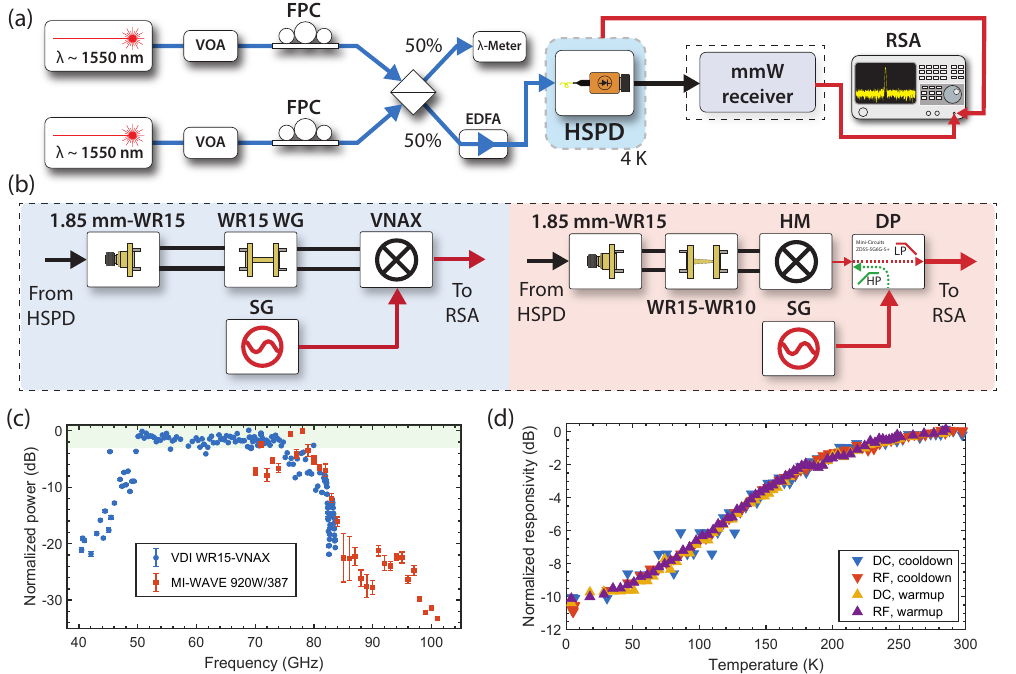}
\caption{\label{fig:HSPD} (a) Schematic of the room-temperature high-speed photodiode measurement setup. The blue lines indicate the optical signal path, the black single lines denote the mm-wave signal path connected by 1.85 mm coaxial cables, the double black lines show mm-wave paths connected by rectangular waveguides, and the red lines indicate the microwave signal path. Note two output paths are coming from the HSPD. The mm-wave signal path is used at room temperature, and the microwave signal path is used when the HSPD is mounted in the cryostat at 4 K. (b) The mm-wave receiver is subdivided into two sections, highlighted in blue and red, corresponding to blue and red data in~\hyperef{Figure}{fig:HSPD} (c). On the left, the mm-wave receiver uses a Virginia Diodes WR15-VNAX to mix down the signal. On the right, the receiver uses an MI-WAVE 920W/387 W-band Harmonic Mixer. (c) Magnitude response of the photodiode as a function of frequency at room temperature. The data are normalized to the maximum measured power. We see a drop in the response at the lower frequencies due to the banded nature of the WR15 VNAX. At higher frequencies, we see a roll-off due to a combination of the HSPD's bandwidth, mixing electronics' response, and insertion loss of coaxial cables. (d) DC and RF responsivity, normalized to their respective room temperature values as a function of temperature.}
\end{figure*}

\clearpage

\begin{thebibliography}{40}%
\makeatletter
\providecommand \@ifxundefined [1]{%
 \@ifx{#1\undefined}
}%
\providecommand \@ifnum [1]{%
 \ifnum #1\expandafter \@firstoftwo
 \else \expandafter \@secondoftwo
 \fi
}%
\providecommand \@ifx [1]{%
 \ifx #1\expandafter \@firstoftwo
 \else \expandafter \@secondoftwo
 \fi
}%
\providecommand \natexlab [1]{#1}%
\providecommand \enquote  [1]{``#1''}%
\providecommand \bibnamefont  [1]{#1}%
\providecommand \bibfnamefont [1]{#1}%
\providecommand \citenamefont [1]{#1}%
\providecommand \href@noop [0]{\@secondoftwo}%
\providecommand \href [0]{\begingroup \@sanitize@url \@href}%
\providecommand \@href[1]{\@@startlink{#1}\@@href}%
\providecommand \@@href[1]{\endgroup#1\@@endlink}%
\providecommand \@sanitize@url [0]{\catcode `\\12\catcode `\$12\catcode `\&12\catcode `\#12\catcode `\^12\catcode `\_12\catcode `\%12\relax}%
\providecommand \@@startlink[1]{}%
\providecommand \@@endlink[0]{}%
\providecommand \url  [0]{\begingroup\@sanitize@url \@url }%
\providecommand \@url [1]{\endgroup\@href {#1}{\urlprefix }}%
\providecommand \urlprefix  [0]{URL }%
\providecommand \Eprint [0]{\href }%
\providecommand \doibase [0]{https://doi.org/}%
\providecommand \selectlanguage [0]{\@gobble}%
\providecommand \bibinfo  [0]{\@secondoftwo}%
\providecommand \bibfield  [0]{\@secondoftwo}%
\providecommand \translation [1]{[#1]}%
\providecommand \BibitemOpen [0]{}%
\providecommand \bibitemStop [0]{}%
\providecommand \bibitemNoStop [0]{.\EOS\space}%
\providecommand \EOS [0]{\spacefactor3000\relax}%
\providecommand \BibitemShut  [1]{\csname bibitem#1\endcsname}%
\let\auto@bib@innerbib\@empty
\bibitem [{\citenamefont {Capmany}\ and\ \citenamefont {Novak}(2007)}]{capmany2007microwave}%
  \BibitemOpen
  \bibfield  {author} {\bibinfo {author} {\bibfnamefont {J.}~\bibnamefont {Capmany}}\ and\ \bibinfo {author} {\bibfnamefont {D.}~\bibnamefont {Novak}},\ }\bibfield  {title} {\bibinfo {title} {Microwave photonics combines two worlds},\ }\href@noop {} {\bibfield  {journal} {\bibinfo  {journal} {Nature photonics}\ }\textbf {\bibinfo {volume} {1}},\ \bibinfo {pages} {319} (\bibinfo {year} {2007})}\BibitemShut {NoStop}%
\bibitem [{\citenamefont {Payne}\ and\ \citenamefont {Shillue}(2002)}]{payne2002photonic}%
  \BibitemOpen
  \bibfield  {author} {\bibinfo {author} {\bibnamefont {Payne}}\ and\ \bibinfo {author} {\bibnamefont {Shillue}},\ }\bibfield  {title} {\bibinfo {title} {Photonic techniques for local oscillator generation and distribution in millimeter-wave radio astronomy},\ }in\ \href@noop {} {\emph {\bibinfo {booktitle} {2002 International Topical Meeting on Microwave Photonics}}}\ (\bibinfo {organization} {IEEE},\ \bibinfo {year} {2002})\ pp.\ \bibinfo {pages} {9--12}\BibitemShut {NoStop}%
\bibitem [{\citenamefont {Bagnell}\ \emph {et~al.}(2013)\citenamefont {Bagnell}, \citenamefont {Davila-Rodriguez},\ and\ \citenamefont {Delfyett}}]{bagnell2013millimeter}%
  \BibitemOpen
  \bibfield  {author} {\bibinfo {author} {\bibfnamefont {M.}~\bibnamefont {Bagnell}}, \bibinfo {author} {\bibfnamefont {J.}~\bibnamefont {Davila-Rodriguez}},\ and\ \bibinfo {author} {\bibfnamefont {P.~J.}\ \bibnamefont {Delfyett}},\ }\bibfield  {title} {\bibinfo {title} {Millimeter-wave generation in an optoelectronic oscillator using an ultrahigh finesse etalon as a photonic filter},\ }\href@noop {} {\bibfield  {journal} {\bibinfo  {journal} {Journal of lightwave technology}\ }\textbf {\bibinfo {volume} {32}},\ \bibinfo {pages} {1063} (\bibinfo {year} {2013})}\BibitemShut {NoStop}%
\bibitem [{\citenamefont {Niu}\ \emph {et~al.}(2015)\citenamefont {Niu}, \citenamefont {Li}, \citenamefont {Jin}, \citenamefont {Su},\ and\ \citenamefont {Vasilakos}}]{Niu2015MoF}%
  \BibitemOpen
  \bibfield  {author} {\bibinfo {author} {\bibfnamefont {Y.}~\bibnamefont {Niu}}, \bibinfo {author} {\bibfnamefont {Y.}~\bibnamefont {Li}}, \bibinfo {author} {\bibfnamefont {D.}~\bibnamefont {Jin}}, \bibinfo {author} {\bibfnamefont {L.}~\bibnamefont {Su}},\ and\ \bibinfo {author} {\bibfnamefont {A.~V.}\ \bibnamefont {Vasilakos}},\ }\bibfield  {title} {\bibinfo {title} {A survey of millimeter wave communications (mmwave) for 5g: opportunities and challenges},\ }\href@noop {} {\bibfield  {journal} {\bibinfo  {journal} {Wireless Networks}\ }\textbf {\bibinfo {volume} {21}},\ \bibinfo {pages} {2657} (\bibinfo {year} {2015})}\BibitemShut {NoStop}%
\bibitem [{\citenamefont {Bardalen}\ \emph {et~al.}(2018)\citenamefont {Bardalen}, \citenamefont {Karlsen}, \citenamefont {Malmbekk}, \citenamefont {Akram},\ and\ \citenamefont {Ohlckers}}]{bardalen2018evaluation}%
  \BibitemOpen
  \bibfield  {author} {\bibinfo {author} {\bibfnamefont {E.}~\bibnamefont {Bardalen}}, \bibinfo {author} {\bibfnamefont {B.}~\bibnamefont {Karlsen}}, \bibinfo {author} {\bibfnamefont {H.}~\bibnamefont {Malmbekk}}, \bibinfo {author} {\bibfnamefont {M.~N.}\ \bibnamefont {Akram}},\ and\ \bibinfo {author} {\bibfnamefont {P.}~\bibnamefont {Ohlckers}},\ }\bibfield  {title} {\bibinfo {title} {Evaluation of {InGaAs/InP} photodiode for high-speed operation at {4 K}},\ }\href@noop {} {\bibfield  {journal} {\bibinfo  {journal} {International Journal of Metrology and Quality Engineering}\ }\textbf {\bibinfo {volume} {9}},\ \bibinfo {pages} {13} (\bibinfo {year} {2018})}\BibitemShut {NoStop}%
\bibitem [{\citenamefont {Burla}\ \emph {et~al.}(2019)\citenamefont {Burla}, \citenamefont {Hoessbacher}, \citenamefont {Heni}, \citenamefont {Haffner}, \citenamefont {Fedoryshyn}, \citenamefont {Werner}, \citenamefont {Watanabe}, \citenamefont {Massler}, \citenamefont {Elder}, \citenamefont {Dalton} \emph {et~al.}}]{burla2019500}%
  \BibitemOpen
  \bibfield  {author} {\bibinfo {author} {\bibfnamefont {M.}~\bibnamefont {Burla}}, \bibinfo {author} {\bibfnamefont {C.}~\bibnamefont {Hoessbacher}}, \bibinfo {author} {\bibfnamefont {W.}~\bibnamefont {Heni}}, \bibinfo {author} {\bibfnamefont {C.}~\bibnamefont {Haffner}}, \bibinfo {author} {\bibfnamefont {Y.}~\bibnamefont {Fedoryshyn}}, \bibinfo {author} {\bibfnamefont {D.}~\bibnamefont {Werner}}, \bibinfo {author} {\bibfnamefont {T.}~\bibnamefont {Watanabe}}, \bibinfo {author} {\bibfnamefont {H.}~\bibnamefont {Massler}}, \bibinfo {author} {\bibfnamefont {D.~L.}\ \bibnamefont {Elder}}, \bibinfo {author} {\bibfnamefont {L.~R.}\ \bibnamefont {Dalton}}, \emph {et~al.},\ }\bibfield  {title} {\bibinfo {title} {500 {GHz} plasmonic mach-zehnder modulator enabling sub-{THz} microwave photonics},\ }\href@noop {} {\bibfield  {journal} {\bibinfo  {journal} {Apl Photonics}\ }\textbf {\bibinfo {volume} {4}} (\bibinfo {year} {2019})}\BibitemShut {NoStop}%
\bibitem [{\citenamefont {Miller}(2017)}]{miller2017attojoule}%
  \BibitemOpen
  \bibfield  {author} {\bibinfo {author} {\bibfnamefont {D.~A.}\ \bibnamefont {Miller}},\ }\bibfield  {title} {\bibinfo {title} {Attojoule optoelectronics for low-energy information processing and communications},\ }\href@noop {} {\bibfield  {journal} {\bibinfo  {journal} {Journal of Lightwave Technology}\ }\textbf {\bibinfo {volume} {35}},\ \bibinfo {pages} {346} (\bibinfo {year} {2017})}\BibitemShut {NoStop}%
\bibitem [{\citenamefont {Marpaung}\ \emph {et~al.}(2019)\citenamefont {Marpaung}, \citenamefont {Yao},\ and\ \citenamefont {Capmany}}]{marpaung2019integrated}%
  \BibitemOpen
  \bibfield  {author} {\bibinfo {author} {\bibfnamefont {D.}~\bibnamefont {Marpaung}}, \bibinfo {author} {\bibfnamefont {J.}~\bibnamefont {Yao}},\ and\ \bibinfo {author} {\bibfnamefont {J.}~\bibnamefont {Capmany}},\ }\bibfield  {title} {\bibinfo {title} {Integrated microwave photonics},\ }\href@noop {} {\bibfield  {journal} {\bibinfo  {journal} {Nature photonics}\ }\textbf {\bibinfo {volume} {13}},\ \bibinfo {pages} {80} (\bibinfo {year} {2019})}\BibitemShut {NoStop}%
\bibitem [{\citenamefont {Zhu}\ \emph {et~al.}(2023)\citenamefont {Zhu}, \citenamefont {Zhang}, \citenamefont {Feng}, \citenamefont {Wang}, \citenamefont {Zhai}, \citenamefont {Feng}, \citenamefont {Pun}, \citenamefont {Zhu},\ and\ \citenamefont {Wang}}]{zhu2023integrated}%
  \BibitemOpen
  \bibfield  {author} {\bibinfo {author} {\bibfnamefont {S.}~\bibnamefont {Zhu}}, \bibinfo {author} {\bibfnamefont {Y.}~\bibnamefont {Zhang}}, \bibinfo {author} {\bibfnamefont {J.}~\bibnamefont {Feng}}, \bibinfo {author} {\bibfnamefont {Y.}~\bibnamefont {Wang}}, \bibinfo {author} {\bibfnamefont {K.}~\bibnamefont {Zhai}}, \bibinfo {author} {\bibfnamefont {H.}~\bibnamefont {Feng}}, \bibinfo {author} {\bibfnamefont {E.~Y.~B.}\ \bibnamefont {Pun}}, \bibinfo {author} {\bibfnamefont {N.~H.}\ \bibnamefont {Zhu}},\ and\ \bibinfo {author} {\bibfnamefont {C.}~\bibnamefont {Wang}},\ }\href@noop {} {\emph {\bibinfo {title} {Integrated lithium niobate photonic millimeter-wave radar}}},\ \bibinfo {type} {Tech. Rep.}\ (\bibinfo {year} {2023})\BibitemShut {NoStop}%
\bibitem [{\citenamefont {Kudelin}\ \emph {et~al.}(2024)\citenamefont {Kudelin}, \citenamefont {Groman}, \citenamefont {Ji}, \citenamefont {Guo}, \citenamefont {Kelleher}, \citenamefont {Lee}, \citenamefont {Nakamura}, \citenamefont {McLemore}, \citenamefont {Shirmohammadi}, \citenamefont {Hanifi} \emph {et~al.}}]{kudelin2024photonic}%
  \BibitemOpen
  \bibfield  {author} {\bibinfo {author} {\bibfnamefont {I.}~\bibnamefont {Kudelin}}, \bibinfo {author} {\bibfnamefont {W.}~\bibnamefont {Groman}}, \bibinfo {author} {\bibfnamefont {Q.-X.}\ \bibnamefont {Ji}}, \bibinfo {author} {\bibfnamefont {J.}~\bibnamefont {Guo}}, \bibinfo {author} {\bibfnamefont {M.~L.}\ \bibnamefont {Kelleher}}, \bibinfo {author} {\bibfnamefont {D.}~\bibnamefont {Lee}}, \bibinfo {author} {\bibfnamefont {T.}~\bibnamefont {Nakamura}}, \bibinfo {author} {\bibfnamefont {C.~A.}\ \bibnamefont {McLemore}}, \bibinfo {author} {\bibfnamefont {P.}~\bibnamefont {Shirmohammadi}}, \bibinfo {author} {\bibfnamefont {S.}~\bibnamefont {Hanifi}}, \emph {et~al.},\ }\bibfield  {title} {\bibinfo {title} {Photonic chip-based low-noise microwave oscillator},\ }\href@noop {} {\bibfield  {journal} {\bibinfo  {journal} {Nature}\ ,\ \bibinfo {pages} {1}} (\bibinfo {year} {2024})}\BibitemShut {NoStop}%
\bibitem [{\citenamefont {Sun}\ \emph {et~al.}(2024)\citenamefont {Sun}, \citenamefont {Wang}, \citenamefont {Liu}, \citenamefont {Harrington}, \citenamefont {Tabatabaei}, \citenamefont {Liu}, \citenamefont {Wang}, \citenamefont {Hanifi}, \citenamefont {Morgan}, \citenamefont {Jahanbozorgi} \emph {et~al.}}]{sun2024integrated}%
  \BibitemOpen
  \bibfield  {author} {\bibinfo {author} {\bibfnamefont {S.}~\bibnamefont {Sun}}, \bibinfo {author} {\bibfnamefont {B.}~\bibnamefont {Wang}}, \bibinfo {author} {\bibfnamefont {K.}~\bibnamefont {Liu}}, \bibinfo {author} {\bibfnamefont {M.~W.}\ \bibnamefont {Harrington}}, \bibinfo {author} {\bibfnamefont {F.}~\bibnamefont {Tabatabaei}}, \bibinfo {author} {\bibfnamefont {R.}~\bibnamefont {Liu}}, \bibinfo {author} {\bibfnamefont {J.}~\bibnamefont {Wang}}, \bibinfo {author} {\bibfnamefont {S.}~\bibnamefont {Hanifi}}, \bibinfo {author} {\bibfnamefont {J.~S.}\ \bibnamefont {Morgan}}, \bibinfo {author} {\bibfnamefont {M.}~\bibnamefont {Jahanbozorgi}}, \emph {et~al.},\ }\bibfield  {title} {\bibinfo {title} {Integrated optical frequency division for microwave and mmwave generation},\ }\href@noop {} {\bibfield  {journal} {\bibinfo  {journal} {Nature}\ ,\ \bibinfo {pages} {1}} (\bibinfo {year} {2024})}\BibitemShut {NoStop}%
\bibitem [{\citenamefont {Lecocq}\ \emph {et~al.}(2021)\citenamefont {Lecocq}, \citenamefont {Quinlan}, \citenamefont {Cicak}, \citenamefont {Aumentado}, \citenamefont {Diddams},\ and\ \citenamefont {Teufel}}]{lecocq2021control}%
  \BibitemOpen
  \bibfield  {author} {\bibinfo {author} {\bibfnamefont {F.}~\bibnamefont {Lecocq}}, \bibinfo {author} {\bibfnamefont {F.}~\bibnamefont {Quinlan}}, \bibinfo {author} {\bibfnamefont {K.}~\bibnamefont {Cicak}}, \bibinfo {author} {\bibfnamefont {J.}~\bibnamefont {Aumentado}}, \bibinfo {author} {\bibfnamefont {S.}~\bibnamefont {Diddams}},\ and\ \bibinfo {author} {\bibfnamefont {J.}~\bibnamefont {Teufel}},\ }\bibfield  {title} {\bibinfo {title} {Control and readout of a superconducting qubit using a photonic link},\ }\href@noop {} {\bibfield  {journal} {\bibinfo  {journal} {Nature}\ }\textbf {\bibinfo {volume} {591}},\ \bibinfo {pages} {575} (\bibinfo {year} {2021})}\BibitemShut {NoStop}%
\bibitem [{\citenamefont {Li}\ \emph {et~al.}(2024)\citenamefont {Li}, \citenamefont {Li}, \citenamefont {Fan}, \citenamefont {Han}, \citenamefont {Xu}, \citenamefont {Lin}, \citenamefont {Guo}, \citenamefont {Li}, \citenamefont {Gong}, \citenamefont {Liao} \emph {et~al.}}]{li2024optical}%
  \BibitemOpen
  \bibfield  {author} {\bibinfo {author} {\bibfnamefont {N.}~\bibnamefont {Li}}, \bibinfo {author} {\bibfnamefont {Y.-H.}\ \bibnamefont {Li}}, \bibinfo {author} {\bibfnamefont {D.-J.}\ \bibnamefont {Fan}}, \bibinfo {author} {\bibfnamefont {L.-C.}\ \bibnamefont {Han}}, \bibinfo {author} {\bibfnamefont {Y.}~\bibnamefont {Xu}}, \bibinfo {author} {\bibfnamefont {J.}~\bibnamefont {Lin}}, \bibinfo {author} {\bibfnamefont {C.}~\bibnamefont {Guo}}, \bibinfo {author} {\bibfnamefont {D.-D.}\ \bibnamefont {Li}}, \bibinfo {author} {\bibfnamefont {M.}~\bibnamefont {Gong}}, \bibinfo {author} {\bibfnamefont {S.-K.}\ \bibnamefont {Liao}}, \emph {et~al.},\ }\bibfield  {title} {\bibinfo {title} {Optical transmission of microwave control signal towards large-scale superconducting quantum computing},\ }\href@noop {} {\bibfield  {journal} {\bibinfo  {journal} {Optics Express}\ }\textbf {\bibinfo {volume} {32}},\ \bibinfo {pages} {3989} (\bibinfo {year} {2024})}\BibitemShut {NoStop}%
\bibitem [{\citenamefont {Faramarzi}\ \emph {et~al.}(2021)\citenamefont {Faramarzi}, \citenamefont {Day}, \citenamefont {Glasby}, \citenamefont {Sypkens}, \citenamefont {Colangelo}, \citenamefont {Chamberlin}, \citenamefont {Mirhosseini}, \citenamefont {Schmidt}, \citenamefont {Berggren},\ and\ \citenamefont {Mauskopf}}]{faramarzi2021initial}%
  \BibitemOpen
  \bibfield  {author} {\bibinfo {author} {\bibfnamefont {F.}~\bibnamefont {Faramarzi}}, \bibinfo {author} {\bibfnamefont {P.}~\bibnamefont {Day}}, \bibinfo {author} {\bibfnamefont {J.}~\bibnamefont {Glasby}}, \bibinfo {author} {\bibfnamefont {S.}~\bibnamefont {Sypkens}}, \bibinfo {author} {\bibfnamefont {M.}~\bibnamefont {Colangelo}}, \bibinfo {author} {\bibfnamefont {R.}~\bibnamefont {Chamberlin}}, \bibinfo {author} {\bibfnamefont {M.}~\bibnamefont {Mirhosseini}}, \bibinfo {author} {\bibfnamefont {K.}~\bibnamefont {Schmidt}}, \bibinfo {author} {\bibfnamefont {K.~K.}\ \bibnamefont {Berggren}},\ and\ \bibinfo {author} {\bibfnamefont {P.}~\bibnamefont {Mauskopf}},\ }\bibfield  {title} {\bibinfo {title} {Initial design of a w-band superconducting kinetic inductance qubit},\ }\href@noop {} {\bibfield  {journal} {\bibinfo  {journal} {IEEE Transactions on Applied Superconductivity}\ }\textbf {\bibinfo {volume} {31}},\ \bibinfo {pages} {1} (\bibinfo {year} {2021})}\BibitemShut {NoStop}%
\bibitem [{\citenamefont {Anferov}\ \emph {et~al.}(2024{\natexlab{a}})\citenamefont {Anferov}, \citenamefont {Lee}, \citenamefont {Zhao}, \citenamefont {Simon},\ and\ \citenamefont {Schuster}}]{anferov2024improved}%
  \BibitemOpen
  \bibfield  {author} {\bibinfo {author} {\bibfnamefont {A.}~\bibnamefont {Anferov}}, \bibinfo {author} {\bibfnamefont {K.-H.}\ \bibnamefont {Lee}}, \bibinfo {author} {\bibfnamefont {F.}~\bibnamefont {Zhao}}, \bibinfo {author} {\bibfnamefont {J.}~\bibnamefont {Simon}},\ and\ \bibinfo {author} {\bibfnamefont {D.~I.}\ \bibnamefont {Schuster}},\ }\bibfield  {title} {\bibinfo {title} {Improved coherence in optically defined niobium trilayer-junction qubits},\ }\href@noop {} {\bibfield  {journal} {\bibinfo  {journal} {Physical Review Applied}\ }\textbf {\bibinfo {volume} {21}},\ \bibinfo {pages} {024047} (\bibinfo {year} {2024}{\natexlab{a}})}\BibitemShut {NoStop}%
\bibitem [{\citenamefont {Anferov}\ \emph {et~al.}(2024{\natexlab{b}})\citenamefont {Anferov}, \citenamefont {Harvey}, \citenamefont {Wan}, \citenamefont {Simon},\ and\ \citenamefont {Schuster}}]{anferov2024superconducting}%
  \BibitemOpen
  \bibfield  {author} {\bibinfo {author} {\bibfnamefont {A.}~\bibnamefont {Anferov}}, \bibinfo {author} {\bibfnamefont {S.~P.}\ \bibnamefont {Harvey}}, \bibinfo {author} {\bibfnamefont {F.}~\bibnamefont {Wan}}, \bibinfo {author} {\bibfnamefont {J.}~\bibnamefont {Simon}},\ and\ \bibinfo {author} {\bibfnamefont {D.~I.}\ \bibnamefont {Schuster}},\ }\href@noop {} {\emph {\bibinfo {title} {Superconducting Qubits Above 20 GHz Operating over 200 mK}}},\ \bibinfo {type} {Tech. Rep.}\ (\bibinfo {year} {2024})\BibitemShut {NoStop}%
\bibitem [{\citenamefont {Stokowski}\ \emph {et~al.}(2019)\citenamefont {Stokowski}, \citenamefont {Pechal}, \citenamefont {Snively}, \citenamefont {Multani}, \citenamefont {Welander}, \citenamefont {Witmer}, \citenamefont {Nanni},\ and\ \citenamefont {Safavi-Naeini}}]{stokowski2019towards}%
  \BibitemOpen
  \bibfield  {author} {\bibinfo {author} {\bibfnamefont {H.}~\bibnamefont {Stokowski}}, \bibinfo {author} {\bibfnamefont {M.}~\bibnamefont {Pechal}}, \bibinfo {author} {\bibfnamefont {E.}~\bibnamefont {Snively}}, \bibinfo {author} {\bibfnamefont {K.~K.}\ \bibnamefont {Multani}}, \bibinfo {author} {\bibfnamefont {P.~B.}\ \bibnamefont {Welander}}, \bibinfo {author} {\bibfnamefont {J.}~\bibnamefont {Witmer}}, \bibinfo {author} {\bibfnamefont {E.~A.}\ \bibnamefont {Nanni}},\ and\ \bibinfo {author} {\bibfnamefont {A.~H.}\ \bibnamefont {Safavi-Naeini}},\ }\bibfield  {title} {\bibinfo {title} {Towards millimeter-wave based quantum networks},\ }in\ \href@noop {} {\emph {\bibinfo {booktitle} {2019 44th International Conference on Infrared, Millimeter, and Terahertz Waves (IRMMW-THz)}}}\ (\bibinfo {organization} {IEEE},\ \bibinfo {year} {2019})\ pp.\ \bibinfo {pages} {1--2}\BibitemShut {NoStop}%
\bibitem [{\citenamefont {Multani}\ \emph {et~al.}(2020)\citenamefont {Multani}, \citenamefont {Stokowski}, \citenamefont {Snively}, \citenamefont {Patel}, \citenamefont {Jiang}, \citenamefont {Lee}, \citenamefont {Welander}, \citenamefont {Nanni},\ and\ \citenamefont {Safavi-Naeini}}]{multani2020development}%
  \BibitemOpen
  \bibfield  {author} {\bibinfo {author} {\bibfnamefont {K.~K.}\ \bibnamefont {Multani}}, \bibinfo {author} {\bibfnamefont {H.}~\bibnamefont {Stokowski}}, \bibinfo {author} {\bibfnamefont {E.}~\bibnamefont {Snively}}, \bibinfo {author} {\bibfnamefont {R.}~\bibnamefont {Patel}}, \bibinfo {author} {\bibfnamefont {W.}~\bibnamefont {Jiang}}, \bibinfo {author} {\bibfnamefont {N.}~\bibnamefont {Lee}}, \bibinfo {author} {\bibfnamefont {P.~B.}\ \bibnamefont {Welander}}, \bibinfo {author} {\bibfnamefont {E.~A.}\ \bibnamefont {Nanni}},\ and\ \bibinfo {author} {\bibfnamefont {A.~H.}\ \bibnamefont {Safavi-Naeini}},\ }\bibfield  {title} {\bibinfo {title} {Development of a millimeter-wave transducer for quantum networks},\ }in\ \href@noop {} {\emph {\bibinfo {booktitle} {2020 45th International Conference on Infrared, Millimeter, and Terahertz Waves (IRMMW-THz)}}}\ (\bibinfo {organization} {IEEE},\ \bibinfo {year} {2020})\ pp.\ \bibinfo {pages} {1--2}\BibitemShut {NoStop}%
\bibitem [{\citenamefont {Anferov}\ \emph {et~al.}(2020)\citenamefont {Anferov}, \citenamefont {Suleymanzade}, \citenamefont {Oriani}, \citenamefont {Simon},\ and\ \citenamefont {Schuster}}]{anferov2020millimeter}%
  \BibitemOpen
  \bibfield  {author} {\bibinfo {author} {\bibfnamefont {A.}~\bibnamefont {Anferov}}, \bibinfo {author} {\bibfnamefont {A.}~\bibnamefont {Suleymanzade}}, \bibinfo {author} {\bibfnamefont {A.}~\bibnamefont {Oriani}}, \bibinfo {author} {\bibfnamefont {J.}~\bibnamefont {Simon}},\ and\ \bibinfo {author} {\bibfnamefont {D.~I.}\ \bibnamefont {Schuster}},\ }\bibfield  {title} {\bibinfo {title} {Millimeter-wave four-wave mixing via kinetic inductance for quantum devices},\ }\href@noop {} {\bibfield  {journal} {\bibinfo  {journal} {Physical Review Applied}\ }\textbf {\bibinfo {volume} {13}},\ \bibinfo {pages} {024056} (\bibinfo {year} {2020})}\BibitemShut {NoStop}%
\bibitem [{\citenamefont {Zhang}\ \emph {et~al.}(1997)\citenamefont {Zhang}, \citenamefont {Borzenets}, \citenamefont {Dubash}, \citenamefont {Reynolds}, \citenamefont {Wey},\ and\ \citenamefont {Bowers}}]{Zhang1997cryogenic}%
  \BibitemOpen
  \bibfield  {author} {\bibinfo {author} {\bibfnamefont {Y.}~\bibnamefont {Zhang}}, \bibinfo {author} {\bibfnamefont {V.}~\bibnamefont {Borzenets}}, \bibinfo {author} {\bibfnamefont {N.}~\bibnamefont {Dubash}}, \bibinfo {author} {\bibfnamefont {T.}~\bibnamefont {Reynolds}}, \bibinfo {author} {\bibfnamefont {Y.}~\bibnamefont {Wey}},\ and\ \bibinfo {author} {\bibfnamefont {J.}~\bibnamefont {Bowers}},\ }\bibfield  {title} {\bibinfo {title} {Cryogenic performance of a high-speed gainas/inp p-i-n photodiode},\ }\href@noop {} {\bibfield  {journal} {\bibinfo  {journal} {Journal of Lightwave Technology}\ }\textbf {\bibinfo {volume} {15}},\ \bibinfo {pages} {529} (\bibinfo {year} {1997})}\BibitemShut {NoStop}%
\bibitem [{\citenamefont {Huggard}\ \emph {et~al.}(2007)\citenamefont {Huggard}, \citenamefont {Fontana}, \citenamefont {Ellison}, \citenamefont {Bortolotti}, \citenamefont {Lazareff},\ and\ \citenamefont {Navarrini}}]{huggard2007photonic}%
  \BibitemOpen
  \bibfield  {author} {\bibinfo {author} {\bibfnamefont {P.}~\bibnamefont {Huggard}}, \bibinfo {author} {\bibfnamefont {A.}~\bibnamefont {Fontana}}, \bibinfo {author} {\bibfnamefont {B.}~\bibnamefont {Ellison}}, \bibinfo {author} {\bibfnamefont {Y.}~\bibnamefont {Bortolotti}}, \bibinfo {author} {\bibfnamefont {B.}~\bibnamefont {Lazareff}},\ and\ \bibinfo {author} {\bibfnamefont {A.}~\bibnamefont {Navarrini}},\ }\bibfield  {title} {\bibinfo {title} {Photonic local oscillator operating at {77 K} for a 2 mm band sis astronomical heterodyne receiver array},\ }in\ \href@noop {} {\emph {\bibinfo {booktitle} {2007 Joint 32nd International Conference on Infrared and Millimeter Waves and the 15th International Conference on Terahertz Electronics}}}\ (\bibinfo {organization} {IEEE},\ \bibinfo {year} {2007})\ pp.\ \bibinfo {pages} {710--711}\BibitemShut {NoStop}%
\bibitem [{\citenamefont {Krantz}\ \emph {et~al.}(2019)\citenamefont {Krantz}, \citenamefont {Kjaergaard}, \citenamefont {Yan}, \citenamefont {Orlando}, \citenamefont {Gustavsson},\ and\ \citenamefont {Oliver}}]{krantz2019quantumeng}%
  \BibitemOpen
  \bibfield  {author} {\bibinfo {author} {\bibfnamefont {P.}~\bibnamefont {Krantz}}, \bibinfo {author} {\bibfnamefont {M.}~\bibnamefont {Kjaergaard}}, \bibinfo {author} {\bibfnamefont {F.}~\bibnamefont {Yan}}, \bibinfo {author} {\bibfnamefont {T.~P.}\ \bibnamefont {Orlando}}, \bibinfo {author} {\bibfnamefont {S.}~\bibnamefont {Gustavsson}},\ and\ \bibinfo {author} {\bibfnamefont {W.~D.}\ \bibnamefont {Oliver}},\ }\bibfield  {title} {\bibinfo {title} {{A quantum engineer's guide to superconducting qubits}},\ }\href@noop {} {\bibfield  {journal} {\bibinfo  {journal} {Applied Physics Reviews}\ }\textbf {\bibinfo {volume} {6}},\ \bibinfo {pages} {021318} (\bibinfo {year} {2019})}\BibitemShut {NoStop}%
\bibitem [{\citenamefont {Clerk}\ \emph {et~al.}(2010)\citenamefont {Clerk}, \citenamefont {Devoret}, \citenamefont {Girvin}, \citenamefont {Marquardt},\ and\ \citenamefont {Schoelkopf}}]{clerk2010introduction}%
  \BibitemOpen
  \bibfield  {author} {\bibinfo {author} {\bibfnamefont {A.~A.}\ \bibnamefont {Clerk}}, \bibinfo {author} {\bibfnamefont {M.~H.}\ \bibnamefont {Devoret}}, \bibinfo {author} {\bibfnamefont {S.~M.}\ \bibnamefont {Girvin}}, \bibinfo {author} {\bibfnamefont {F.}~\bibnamefont {Marquardt}},\ and\ \bibinfo {author} {\bibfnamefont {R.~J.}\ \bibnamefont {Schoelkopf}},\ }\bibfield  {title} {\bibinfo {title} {Introduction to quantum noise, measurement, and amplification},\ }\href@noop {} {\bibfield  {journal} {\bibinfo  {journal} {Reviews of Modern Physics}\ }\textbf {\bibinfo {volume} {82}},\ \bibinfo {pages} {1155} (\bibinfo {year} {2010})}\BibitemShut {NoStop}%
\bibitem [{\citenamefont {Krinner}\ \emph {et~al.}(2019)\citenamefont {Krinner}, \citenamefont {Storz}, \citenamefont {Kurpiers}, \citenamefont {Magnard}, \citenamefont {Heinsoo}, \citenamefont {Keller}, \citenamefont {Luetolf}, \citenamefont {Eichler},\ and\ \citenamefont {Wallraff}}]{krinner2019engineering}%
  \BibitemOpen
  \bibfield  {author} {\bibinfo {author} {\bibfnamefont {S.}~\bibnamefont {Krinner}}, \bibinfo {author} {\bibfnamefont {S.}~\bibnamefont {Storz}}, \bibinfo {author} {\bibfnamefont {P.}~\bibnamefont {Kurpiers}}, \bibinfo {author} {\bibfnamefont {P.}~\bibnamefont {Magnard}}, \bibinfo {author} {\bibfnamefont {J.}~\bibnamefont {Heinsoo}}, \bibinfo {author} {\bibfnamefont {R.}~\bibnamefont {Keller}}, \bibinfo {author} {\bibfnamefont {J.}~\bibnamefont {Luetolf}}, \bibinfo {author} {\bibfnamefont {C.}~\bibnamefont {Eichler}},\ and\ \bibinfo {author} {\bibfnamefont {A.}~\bibnamefont {Wallraff}},\ }\bibfield  {title} {\bibinfo {title} {Engineering cryogenic setups for 100-qubit scale superconducting circuit systems},\ }\href@noop {} {\bibfield  {journal} {\bibinfo  {journal} {EPJ Quantum Technology}\ }\textbf {\bibinfo {volume} {6}},\ \bibinfo {pages} {2} (\bibinfo {year} {2019})}\BibitemShut {NoStop}%
\bibitem [{\citenamefont {Meesala}\ \emph {et~al.}(2024)\citenamefont {Meesala}, \citenamefont {Wood}, \citenamefont {Lake}, \citenamefont {Chiappina}, \citenamefont {Zhong}, \citenamefont {Beyer}, \citenamefont {Shaw}, \citenamefont {Jiang},\ and\ \citenamefont {Painter}}]{meesala2024non}%
  \BibitemOpen
  \bibfield  {author} {\bibinfo {author} {\bibfnamefont {S.}~\bibnamefont {Meesala}}, \bibinfo {author} {\bibfnamefont {S.}~\bibnamefont {Wood}}, \bibinfo {author} {\bibfnamefont {D.}~\bibnamefont {Lake}}, \bibinfo {author} {\bibfnamefont {P.}~\bibnamefont {Chiappina}}, \bibinfo {author} {\bibfnamefont {C.}~\bibnamefont {Zhong}}, \bibinfo {author} {\bibfnamefont {A.~D.}\ \bibnamefont {Beyer}}, \bibinfo {author} {\bibfnamefont {M.~D.}\ \bibnamefont {Shaw}}, \bibinfo {author} {\bibfnamefont {L.}~\bibnamefont {Jiang}},\ and\ \bibinfo {author} {\bibfnamefont {O.}~\bibnamefont {Painter}},\ }\bibfield  {title} {\bibinfo {title} {Non-classical microwave--optical photon pair generation with a chip-scale transducer},\ }\href@noop {} {\bibfield  {journal} {\bibinfo  {journal} {Nature Physics}\ ,\ \bibinfo {pages} {1}} (\bibinfo {year} {2024})}\BibitemShut {NoStop}%
\bibitem [{\citenamefont {Jiang}\ \emph {et~al.}(2023)\citenamefont {Jiang}, \citenamefont {Mayor}, \citenamefont {Malik}, \citenamefont {Van~Laer}, \citenamefont {McKenna}, \citenamefont {Patel}, \citenamefont {Witmer},\ and\ \citenamefont {Safavi-Naeini}}]{jiang2023optically}%
  \BibitemOpen
  \bibfield  {author} {\bibinfo {author} {\bibfnamefont {W.}~\bibnamefont {Jiang}}, \bibinfo {author} {\bibfnamefont {F.~M.}\ \bibnamefont {Mayor}}, \bibinfo {author} {\bibfnamefont {S.}~\bibnamefont {Malik}}, \bibinfo {author} {\bibfnamefont {R.}~\bibnamefont {Van~Laer}}, \bibinfo {author} {\bibfnamefont {T.~P.}\ \bibnamefont {McKenna}}, \bibinfo {author} {\bibfnamefont {R.~N.}\ \bibnamefont {Patel}}, \bibinfo {author} {\bibfnamefont {J.~D.}\ \bibnamefont {Witmer}},\ and\ \bibinfo {author} {\bibfnamefont {A.~H.}\ \bibnamefont {Safavi-Naeini}},\ }\bibfield  {title} {\bibinfo {title} {Optically heralded microwave photon addition},\ }\href@noop {} {\bibfield  {journal} {\bibinfo  {journal} {Nature Physics}\ }\textbf {\bibinfo {volume} {19}},\ \bibinfo {pages} {1423} (\bibinfo {year} {2023})}\BibitemShut {NoStop}%
\bibitem [{\citenamefont {Delaney}\ \emph {et~al.}(2022)\citenamefont {Delaney}, \citenamefont {Urmey}, \citenamefont {Mittal}, \citenamefont {Brubaker}, \citenamefont {Kindem}, \citenamefont {Burns}, \citenamefont {Regal},\ and\ \citenamefont {Lehnert}}]{delaney2022superconducting}%
  \BibitemOpen
  \bibfield  {author} {\bibinfo {author} {\bibfnamefont {R.}~\bibnamefont {Delaney}}, \bibinfo {author} {\bibfnamefont {M.}~\bibnamefont {Urmey}}, \bibinfo {author} {\bibfnamefont {S.}~\bibnamefont {Mittal}}, \bibinfo {author} {\bibfnamefont {B.}~\bibnamefont {Brubaker}}, \bibinfo {author} {\bibfnamefont {J.}~\bibnamefont {Kindem}}, \bibinfo {author} {\bibfnamefont {P.}~\bibnamefont {Burns}}, \bibinfo {author} {\bibfnamefont {C.}~\bibnamefont {Regal}},\ and\ \bibinfo {author} {\bibfnamefont {K.}~\bibnamefont {Lehnert}},\ }\bibfield  {title} {\bibinfo {title} {Superconducting-qubit readout via low-backaction electro-optic transduction},\ }\href@noop {} {\bibfield  {journal} {\bibinfo  {journal} {Nature}\ }\textbf {\bibinfo {volume} {606}},\ \bibinfo {pages} {489} (\bibinfo {year} {2022})}\BibitemShut {NoStop}%
\bibitem [{\citenamefont {Han}\ \emph {et~al.}(2021)\citenamefont {Han}, \citenamefont {Fu}, \citenamefont {Zou}, \citenamefont {Jiang},\ and\ \citenamefont {Tang}}]{han2021microwave}%
  \BibitemOpen
  \bibfield  {author} {\bibinfo {author} {\bibfnamefont {X.}~\bibnamefont {Han}}, \bibinfo {author} {\bibfnamefont {W.}~\bibnamefont {Fu}}, \bibinfo {author} {\bibfnamefont {C.-L.}\ \bibnamefont {Zou}}, \bibinfo {author} {\bibfnamefont {L.}~\bibnamefont {Jiang}},\ and\ \bibinfo {author} {\bibfnamefont {H.~X.}\ \bibnamefont {Tang}},\ }\bibfield  {title} {\bibinfo {title} {Microwave-optical quantum frequency conversion},\ }\href@noop {} {\bibfield  {journal} {\bibinfo  {journal} {Optica}\ }\textbf {\bibinfo {volume} {8}},\ \bibinfo {pages} {1050} (\bibinfo {year} {2021})}\BibitemShut {NoStop}%
\bibitem [{\citenamefont {Lambert}\ \emph {et~al.}(2020)\citenamefont {Lambert}, \citenamefont {Rueda}, \citenamefont {Sedlmeir},\ and\ \citenamefont {Schwefel}}]{lambert2020coherent}%
  \BibitemOpen
  \bibfield  {author} {\bibinfo {author} {\bibfnamefont {N.~J.}\ \bibnamefont {Lambert}}, \bibinfo {author} {\bibfnamefont {A.}~\bibnamefont {Rueda}}, \bibinfo {author} {\bibfnamefont {F.}~\bibnamefont {Sedlmeir}},\ and\ \bibinfo {author} {\bibfnamefont {H.~G.}\ \bibnamefont {Schwefel}},\ }\bibfield  {title} {\bibinfo {title} {Coherent conversion between microwave and optical photons—an overview of physical implementations},\ }\href@noop {} {\bibfield  {journal} {\bibinfo  {journal} {Advanced Quantum Technologies}\ }\textbf {\bibinfo {volume} {3}},\ \bibinfo {pages} {1900077} (\bibinfo {year} {2020})}\BibitemShut {NoStop}%
\bibitem [{\citenamefont {Brubaker}\ \emph {et~al.}(2022)\citenamefont {Brubaker}, \citenamefont {Kindem}, \citenamefont {Urmey}, \citenamefont {Mittal}, \citenamefont {Delaney}, \citenamefont {Burns}, \citenamefont {Vissers}, \citenamefont {Lehnert},\ and\ \citenamefont {Regal}}]{brubaker2022optomechanical}%
  \BibitemOpen
  \bibfield  {author} {\bibinfo {author} {\bibfnamefont {B.~M.}\ \bibnamefont {Brubaker}}, \bibinfo {author} {\bibfnamefont {J.~M.}\ \bibnamefont {Kindem}}, \bibinfo {author} {\bibfnamefont {M.~D.}\ \bibnamefont {Urmey}}, \bibinfo {author} {\bibfnamefont {S.}~\bibnamefont {Mittal}}, \bibinfo {author} {\bibfnamefont {R.~D.}\ \bibnamefont {Delaney}}, \bibinfo {author} {\bibfnamefont {P.~S.}\ \bibnamefont {Burns}}, \bibinfo {author} {\bibfnamefont {M.~R.}\ \bibnamefont {Vissers}}, \bibinfo {author} {\bibfnamefont {K.~W.}\ \bibnamefont {Lehnert}},\ and\ \bibinfo {author} {\bibfnamefont {C.~A.}\ \bibnamefont {Regal}},\ }\bibfield  {title} {\bibinfo {title} {Optomechanical ground-state cooling in a continuous and efficient electro-optic transducer},\ }\href@noop {} {\bibfield  {journal} {\bibinfo  {journal} {Physical Review X}\ }\textbf {\bibinfo {volume} {12}},\ \bibinfo {pages} {021062} (\bibinfo {year} {2022})}\BibitemShut {NoStop}%
\bibitem [{\citenamefont {Tu}\ \emph {et~al.}(2022)\citenamefont {Tu}, \citenamefont {Liao}, \citenamefont {Zhang}, \citenamefont {Liu}, \citenamefont {Zheng}, \citenamefont {Yang}, \citenamefont {Zhang}, \citenamefont {Yan},\ and\ \citenamefont {Zhu}}]{tu2022high}%
  \BibitemOpen
  \bibfield  {author} {\bibinfo {author} {\bibfnamefont {H.-T.}\ \bibnamefont {Tu}}, \bibinfo {author} {\bibfnamefont {K.-Y.}\ \bibnamefont {Liao}}, \bibinfo {author} {\bibfnamefont {Z.-X.}\ \bibnamefont {Zhang}}, \bibinfo {author} {\bibfnamefont {X.-H.}\ \bibnamefont {Liu}}, \bibinfo {author} {\bibfnamefont {S.-Y.}\ \bibnamefont {Zheng}}, \bibinfo {author} {\bibfnamefont {S.-Z.}\ \bibnamefont {Yang}}, \bibinfo {author} {\bibfnamefont {X.-D.}\ \bibnamefont {Zhang}}, \bibinfo {author} {\bibfnamefont {H.}~\bibnamefont {Yan}},\ and\ \bibinfo {author} {\bibfnamefont {S.-L.}\ \bibnamefont {Zhu}},\ }\bibfield  {title} {\bibinfo {title} {High-efficiency coherent microwave-to-optics conversion via off-resonant scattering},\ }\href@noop {} {\bibfield  {journal} {\bibinfo  {journal} {Nature Photonics}\ }\textbf {\bibinfo {volume} {16}},\ \bibinfo {pages} {291} (\bibinfo {year} {2022})}\BibitemShut {NoStop}%
\bibitem [{\citenamefont {Sahu}\ \emph {et~al.}(2022)\citenamefont {Sahu}, \citenamefont {Hease}, \citenamefont {Rueda}, \citenamefont {Arnold}, \citenamefont {Qiu},\ and\ \citenamefont {Fink}}]{sahu2022quantum}%
  \BibitemOpen
  \bibfield  {author} {\bibinfo {author} {\bibfnamefont {R.}~\bibnamefont {Sahu}}, \bibinfo {author} {\bibfnamefont {W.}~\bibnamefont {Hease}}, \bibinfo {author} {\bibfnamefont {A.}~\bibnamefont {Rueda}}, \bibinfo {author} {\bibfnamefont {G.}~\bibnamefont {Arnold}}, \bibinfo {author} {\bibfnamefont {L.}~\bibnamefont {Qiu}},\ and\ \bibinfo {author} {\bibfnamefont {J.~M.}\ \bibnamefont {Fink}},\ }\bibfield  {title} {\bibinfo {title} {Quantum-enabled operation of a microwave-optical interface},\ }\href@noop {} {\bibfield  {journal} {\bibinfo  {journal} {Nature communications}\ }\textbf {\bibinfo {volume} {13}},\ \bibinfo {pages} {1276} (\bibinfo {year} {2022})}\BibitemShut {NoStop}%
\bibitem [{\citenamefont {Wikus}\ and\ \citenamefont {Niinikoski}(2010)}]{wikus2010theoretical}%
  \BibitemOpen
  \bibfield  {author} {\bibinfo {author} {\bibfnamefont {P.}~\bibnamefont {Wikus}}\ and\ \bibinfo {author} {\bibfnamefont {T.~O.}\ \bibnamefont {Niinikoski}},\ }\bibfield  {title} {\bibinfo {title} {Theoretical models for the cooling power and base temperature of dilution refrigerators},\ }\href@noop {} {\bibfield  {journal} {\bibinfo  {journal} {Journal of Low Temperature Physics}\ }\textbf {\bibinfo {volume} {158}},\ \bibinfo {pages} {901} (\bibinfo {year} {2010})}\BibitemShut {NoStop}%
\bibitem [{\citenamefont {Setiawan~Putra}\ \emph {et~al.}(2017)\citenamefont {Setiawan~Putra}, \citenamefont {Yamada}, \citenamefont {Tsuda},\ and\ \citenamefont {Ambran}}]{putra2017edfanoise}%
  \BibitemOpen
  \bibfield  {author} {\bibinfo {author} {\bibfnamefont {A.~W.}\ \bibnamefont {Setiawan~Putra}}, \bibinfo {author} {\bibfnamefont {M.}~\bibnamefont {Yamada}}, \bibinfo {author} {\bibfnamefont {H.}~\bibnamefont {Tsuda}},\ and\ \bibinfo {author} {\bibfnamefont {S.}~\bibnamefont {Ambran}},\ }\bibfield  {title} {\bibinfo {title} {Theoretical analysis of noise in erbium doped fiber amplifier},\ }\href@noop {} {\bibfield  {journal} {\bibinfo  {journal} {IEEE Journal of Quantum Electronics}\ }\textbf {\bibinfo {volume} {53}},\ \bibinfo {pages} {1} (\bibinfo {year} {2017})}\BibitemShut {NoStop}%
\bibitem [{\citenamefont {Pechal}\ and\ \citenamefont {Safavi-Naeini}(2017)}]{pechal2017millimeter}%
  \BibitemOpen
  \bibfield  {author} {\bibinfo {author} {\bibfnamefont {M.}~\bibnamefont {Pechal}}\ and\ \bibinfo {author} {\bibfnamefont {A.~H.}\ \bibnamefont {Safavi-Naeini}},\ }\bibfield  {title} {\bibinfo {title} {Millimeter-wave interconnects for microwave-frequency quantum machines},\ }\href@noop {} {\bibfield  {journal} {\bibinfo  {journal} {Physical Review A}\ }\textbf {\bibinfo {volume} {96}},\ \bibinfo {pages} {042305} (\bibinfo {year} {2017})}\BibitemShut {NoStop}%
\bibitem [{\citenamefont {Kumar}\ \emph {et~al.}(2023)\citenamefont {Kumar}, \citenamefont {Suleymanzade}, \citenamefont {Stone}, \citenamefont {Taneja}, \citenamefont {Anferov}, \citenamefont {Schuster},\ and\ \citenamefont {Simon}}]{kumar2023quantum}%
  \BibitemOpen
  \bibfield  {author} {\bibinfo {author} {\bibfnamefont {A.}~\bibnamefont {Kumar}}, \bibinfo {author} {\bibfnamefont {A.}~\bibnamefont {Suleymanzade}}, \bibinfo {author} {\bibfnamefont {M.}~\bibnamefont {Stone}}, \bibinfo {author} {\bibfnamefont {L.}~\bibnamefont {Taneja}}, \bibinfo {author} {\bibfnamefont {A.}~\bibnamefont {Anferov}}, \bibinfo {author} {\bibfnamefont {D.~I.}\ \bibnamefont {Schuster}},\ and\ \bibinfo {author} {\bibfnamefont {J.}~\bibnamefont {Simon}},\ }\bibfield  {title} {\bibinfo {title} {Quantum-enabled millimetre wave to optical transduction using neutral atoms},\ }\href@noop {} {\bibfield  {journal} {\bibinfo  {journal} {Nature}\ }\textbf {\bibinfo {volume} {615}},\ \bibinfo {pages} {614} (\bibinfo {year} {2023})}\BibitemShut {NoStop}%
\bibitem [{\citenamefont {Tan}\ \emph {et~al.}(2024)\citenamefont {Tan}, \citenamefont {Klimovich}, \citenamefont {Stephenson}, \citenamefont {Faramarzi},\ and\ \citenamefont {Day}}]{tan2024operation}%
  \BibitemOpen
  \bibfield  {author} {\bibinfo {author} {\bibfnamefont {B.-K.}\ \bibnamefont {Tan}}, \bibinfo {author} {\bibfnamefont {N.}~\bibnamefont {Klimovich}}, \bibinfo {author} {\bibfnamefont {R.}~\bibnamefont {Stephenson}}, \bibinfo {author} {\bibfnamefont {F.}~\bibnamefont {Faramarzi}},\ and\ \bibinfo {author} {\bibfnamefont {P.}~\bibnamefont {Day}},\ }\bibfield  {title} {\bibinfo {title} {Operation of kinetic-inductance travelling wave parametric amplifiers at millimetre wavelengths},\ }\href@noop {} {\bibfield  {journal} {\bibinfo  {journal} {Superconductor Science and Technology}\ }\textbf {\bibinfo {volume} {37}},\ \bibinfo {pages} {035006} (\bibinfo {year} {2024})}\BibitemShut {NoStop}%
\bibitem [{\citenamefont {Vaartjes}\ \emph {et~al.}(2023)\citenamefont {Vaartjes}, \citenamefont {Kringh{\o}j}, \citenamefont {Vine}, \citenamefont {Day}, \citenamefont {Morello},\ and\ \citenamefont {Pla}}]{vaartjes2023strong}%
  \BibitemOpen
  \bibfield  {author} {\bibinfo {author} {\bibfnamefont {A.}~\bibnamefont {Vaartjes}}, \bibinfo {author} {\bibfnamefont {A.}~\bibnamefont {Kringh{\o}j}}, \bibinfo {author} {\bibfnamefont {W.}~\bibnamefont {Vine}}, \bibinfo {author} {\bibfnamefont {T.}~\bibnamefont {Day}}, \bibinfo {author} {\bibfnamefont {A.}~\bibnamefont {Morello}},\ and\ \bibinfo {author} {\bibfnamefont {J.~J.}\ \bibnamefont {Pla}},\ }\href@noop {} {\emph {\bibinfo {title} {Strong Microwave Squeezing Above 1 Tesla and 1 Kelvin}}},\ \bibinfo {type} {Tech. Rep.}\ (\bibinfo {year} {2023})\BibitemShut {NoStop}%
\bibitem [{\citenamefont {Blais}\ \emph {et~al.}(2004)\citenamefont {Blais}, \citenamefont {Huang}, \citenamefont {Wallraff}, \citenamefont {Girvin},\ and\ \citenamefont {Schoelkopf}}]{blais2004circuitqed}%
  \BibitemOpen
  \bibfield  {author} {\bibinfo {author} {\bibfnamefont {A.}~\bibnamefont {Blais}}, \bibinfo {author} {\bibfnamefont {R.-S.}\ \bibnamefont {Huang}}, \bibinfo {author} {\bibfnamefont {A.}~\bibnamefont {Wallraff}}, \bibinfo {author} {\bibfnamefont {S.~M.}\ \bibnamefont {Girvin}},\ and\ \bibinfo {author} {\bibfnamefont {R.~J.}\ \bibnamefont {Schoelkopf}},\ }\bibfield  {title} {\bibinfo {title} {Cavity quantum electrodynamics for superconducting electrical circuits: An architecture for quantum computation},\ }\href@noop {} {\bibfield  {journal} {\bibinfo  {journal} {Phys. Rev. A}\ }\textbf {\bibinfo {volume} {69}},\ \bibinfo {pages} {062320} (\bibinfo {year} {2004})}\BibitemShut {NoStop}%
\bibitem [{Note1()}]{Note1}%
  \BibitemOpen
  \bibinfo {note} {In comparison, the frequency extender has a larger footprint and higher power demand (9 V/45 W, 3 in x 13 in x 3 in).}\BibitemShut {Stop}%
\end{thebibliography}
%

\end{document}